\documentclass[a4paper,11pt]{article}
\pdfoutput=1
\usepackage{jcappub}

\usepackage{tikz,xcolor,hyperref}
\usepackage{amsmath, amssymb, amsthm, graphicx, epsfig, fancyhdr,epsfig, slashed}
\usepackage[normalem]{ulem}
\usepackage{tikzsymbols}
\usepackage{natbib}
\usepackage{float}
\newcommand{\gs}{g_\star}
\newcommand{\gss}{g_{\star s}}
\newcommand{\Trh}{T_\text{rh}}
\newcommand{\arh}{a_\text{rh}}
\newcommand{\Tmax}{T_\text{max}}

\newcommand{\rGW}{\rho_\text{GW}}
\newcommand{\rR}{\rho_R}
\newcommand{\rp}{\rho_\phi}

\newcommand{\Gp}{\Gamma_\phi}

\newcommand{\mueff}{\mu_\text{eff}}
\newcommand{\yeff}{y_\text{eff}}
\newcommand{\ndm}{n_{\rm dm}}
\newcommand{\Ndm}{N_{\rm dm}}
\newcommand{\mdm}{m_{\rm dm}}
\newcommand{\adm}{a_{\rm dm}}
\newcommand{\lNP}{\Lambda_{\rm NP}}

\newcommand{\ogw}{\Omega_\text{GW}}
\newcommand{\ahc}{a_\text{hc}}
\newcommand{\aeq}{a_{\text{eq}}}
\newcommand{\frh}{f_\text{rh}}
\newcommand{\fmax}{f_\text{max}}

\newcommand{\Thc}{T_\text{hc}}
\newcommand{\Teq}{T_\text{eq}}
\newcommand{\meff}{\mu_\text{eff}}
\newcommand{\yNN}{y_{\phi N}}

\begin{document}
\title{Testing leptogenesis and dark matter production during reheating with primordial gravitational waves }
\author[a]{Basabendu Barman,}
\author[a]{Arindam Basu,}
\author[b]{Debasish Borah,}
\author[a]{Amit Chakraborty}
\author[c]{and Rishav Roshan}
\affiliation[a]{\,\,Department of Physics, School of Engineering and Sciences, SRM University AP, Amaravati 522240, India}
\affiliation[b]{\,\,Department of Physics, Indian Institute of Technology Guwahati, Assam 781039, India}
\affiliation[c]{\,\,School of Physics and Astronomy, University of Southampton,
Southampton, United Kingdom}
\emailAdd{basabendu.b@srmap.edu.in}
\emailAdd{arindam\_basu@srmap.edu.in}
\emailAdd{amit.c@srmap.edu.in}
\emailAdd{dborah@iitg.ac.in}
\emailAdd{r.roshan@soton.ac.uk}
\abstract{We study the generation of of  baryon asymmetry as well as dark matter (DM) in an extended reheating period after the end of slow-roll inflation. Within the regime of perturbative reheating, we consider different monomial potential of the inflaton field during reheating era. The inflaton condensate reheats the Universe by decaying into the Standard Model (SM) bath either via fermionic or bosonic decay modes. Assuming the leptogenesis route to baryogenesis in a canonical seesaw framework, we consider both the radiation bath and perturbative inflaton decay to produce such RHNs during the period of reheating when the maximum temperature of the SM bath is well above the reheating temperature. The DM, assumed to be a SM gauge singlet field, also gets produced from the bath during the reheating period via UV freeze-in. In addition to obtaining different parameter space for such non-thermal leptogenesis and DM for both bosonic and fermionic reheating modes and the type of monomial potential, we discuss the possibility of probing such scenarios via spectral shape of primordial gravitational waves.}
\maketitle
\section{Introduction}
\label{sec:intro}
Observations from two distinct cosmological measurements—the cosmic microwave background (CMB)~\cite{Planck:2018vyg} and big bang nucleosynthesis (BBN)~\cite{ParticleDataGroup:2020ssz}—both confirm that the visible (baryonic) matter in the Universe is asymmetric. This baryon asymmetry presents a significant puzzle in particle physics. According to Sakharov’s conditions~\cite{Sakharov:1967dj}, three essential criteria are required to be satisfied for dynamical generation of this asymmetry: (i) violation of baryon number, (ii) violation of $C$ and $CP$ symmetries, and (iii) out-of-equilibrium dynamics. Although the Standard Model (SM) of particle physics, in principle, contains these elements, it cannot account for the magnitude of the observed asymmetry. To address this, one must look beyond the SM. A particularly compelling mechanism to generate the baryon asymmetry of the Universe (BAU) through the physics of the lepton sector is known as leptogenesis~\cite{Fukugita:1986hr}. In this mechanism, instead of directly creating a baryon asymmetry, a lepton asymmetry is first produced and then converted into baryon asymmetry via $(B+L)$-violating electroweak sphaleron transitions~\cite{Kuzmin:1985mm}. If leptogenesis is implemented in Type-I seesaw framework, also explaining the origin of neutrino mass~\cite{Minkowski:1977sc,GellMann:1980vs,Mohapatra:1979ia,Schechter:1980gr,Schechter:1981cv}, the complex Yukawa couplings of right-handed neutrinos (RHNs) with SM lepton and Higgs doublets, provide the necessary source of CP violation, while the Universe’s expansion (captured by the Hubble rate) ensures a departure from the thermal equilibrium of the decaying RHN. Lepton number violation arises from the Majorana masses of these new RHN fields. In standard thermal leptogenesis (for a review, see Ref.~\cite{Davidson:2008bu}), RHNs are thermally produced from the SM plasma. However, a lower bound on the RHN mass (known as the Davidson-Ibarra bound) sets a constraint on the reheating temperature, requiring $\Trh\gtrsim 10^{10}$ GeV~\cite{Davidson:2002qv}. Such tight constraints can be relaxed if the RHNs are produced in the reheating era itself \cite{Lazarides:1991wu,Murayama:1992ua,Kolb:1996jt,Giudice:1999fb,Asaka:1999yd,Asaka:1999jb,Hamaguchi:2001gw,Hahn-Woernle:2008tsk,Hamada:2018epb,Eijima:2019hey,Maleknejad:2020pec,Barman:2021tgt,Barman:2021ost,Barman:2022gjo,Lazarides:2022ezc,Datta:2022jic,Datta:2023pav,Barman:2024jqh}\footnote{In Ref.~\cite{Flores:2024lzv}, the authors have discussed generation of baryon asymmetry via leptogenesis considering gravitational particle production, employing the Bogoliubov approach. Ref.~\cite{Domcke:2020quw} discusses ``wash-in" leptogenesis, where the RHNs can have masses as low as a few 100 TeV.}. Once produced, they further decay to generate the matter-antimatter asymmetry of the Universe.

While baryonic matter accounts for only about 4\% of the total matter-energy content of the Universe~\cite{Planck:2018vyg}, the 26\% is attributed to dark matter (DM)~\cite{Jungman:1995df,Bertone:2016nfn, deSwart:2017heh}, one of the most profound and unresolved mysteries in both particle physics and cosmology. Tight observational constraints, particularly from direct-detection experiments like the LUX-ZEPLIN (LZ) \cite{LZ:2022lsv}, on the parameter space of traditional weakly interacting massive particles (WIMPs), which are leading DM candidates (see, e.g., Refs.~\cite{Roszkowski:2017nbc, Arcadi:2017kky}), have motivated the exploration of alternative DM production mechanisms. One popular alternative among these is the concept of feebly interacting massive particles (FIMPs), which can be produced in the early Universe via the decay or annihilation of particles in the visible sector. Once the temperature of the SM plasma falls below the characteristic interaction mass scale, DM production becomes Boltzmann suppressed. This results in a constant comoving number density of DM, thus realizing what is known as freeze-in~\cite{McDonald:2001vt, Hall:2009bx, Bernal:2017kxu, Datta:2021elq}. The FIMP scenario requires extremely suppressed interactions between the dark and visible sectors to ensure the non-thermal production of DM. Such interactions may be achieved either through tiny couplings in the infrared (IR) regime, or via non-renormalizable operators suppressed by a high-energy new physics (NP) scale in the case of ultraviolet (UV) freeze-in~\cite{Hall:2009bx, Elahi:2014fsa,Barman:2020plp}. The UV freeze-in scenario is particularly intriguing because the DM yield depends on the maximum temperature, $\Tmax$, reached by the SM plasma~\cite{Giudice:2000ex, Garcia:2020eof}, which, in turn, is governed by the post-inflationary dynamics of the inflaton field.

The inflationary gravitational wave (GW) is one of the most vital predictions of the inflationary paradigm (for details, see for example,  Refs.~\cite{Grishchuk:1974ny,Starobinsky:1979ty,Guzzetti:2016mkm,Roshan:2024qnv}). During inflation, quantum fluctuations inevitably generate a scale-invariant spectrum of tensor metric perturbations on super-Hubble scales. In the standard post-inflationary scenario, these tensor modes re-enter the horizon during the radiation-dominated (RD) era. However, it is well known that the introduction of non-standard cosmologies—such as an early stiff epoch preceding radiation domination—breaks this scale invariance. Under such circumstances, the GW spectrum acquires a pronounced blue tilt in the frequency range corresponding to modes that re-entered the horizon during the stiff era~\cite{Giovannini:1998bp,Giovannini:1999bh,Riazuelo:2000fc,Seto:2003kc,Boyle:2007zx,Stewart:2007fu,Li:2021htg,Artymowski:2017pua,Caprini:2018mtu,Bettoni:2018pbl,Figueroa:2019paj,Opferkuch:2019zbd,Bernal:2020ywq,Ghoshal:2022ruy,Caldwell:2022qsj,Gouttenoire:2021jhk,Haque:2021dha,Chakraborty:2023ocr,Barman:2022qgt,Barman:2023ktz,Barman:2024slw,Barman:2024mqo}. As discussed in~\cite{Figueroa:2019paj,Ghoshal:2023phi,Barman:2023ktz}, although a stiff period in the expansion history enhances significantly the inflationary GW background, making this signal potentially observable, however it may also lead to over production of GW energy density, thereby violating standard BBN and CMB bounds on GW backgrounds. 

Motivated from these, in this work we have investigated dynamical generation of baryon asymmetry via leptogenesis, by considering CP-violating decay of the RHNs originated from the radiation bath, as well as from the perturbative inflaton decay during the period of reheating. We consider reheating to take place from the decay of the inflaton either into a pair of SM-like fermions or into a pair of SM-like bosons, while the inflaton oscillates at the bottom of a monomial potential. We also consider DM production from the thermal bath during reheating through non-renormalizable interactions, that gives rise to UV freeze-in, where bulk of the DM production occurs around the highest temperature of the thermal bath. Since production of both the DM and baryon asymmetry is controlled by the background dynamics, namely, inflaton-SM coupling and shape of the inflaton potential during reheating, we check the feasibility of simultaneously producing both in the required amount for certain choices of the model parameters. It is also possible to test our scenario at several, futuristic GW detection facilities, thanks to the blue-tilt imparted to the primordial GW spectrum by the stiff expansion history of the reheating era. It is worth mentioning that earlier studies, for example, Refs.~\cite{Hahn-Woernle:2008tsk,Barman:2021tgt,Datta:2022jic} did consider leptogenesis during reheating, but for quadratic inflaton potential, whereas Ref.~\cite{Hahn-Woernle:2008tsk,Barman:2021tgt,Datta:2023pav} also considered explicit coupling between inflaton and the RHNs. The present study differs from the previous ones in two ways: (a) inclusion of generic monomial potential and (b) having primordial GW as a probe for the scale of leptogenesis. In addition, we also explore production of dark matter via UV freeze-in from the radiation bath during reheating. However, we always consider the inflaton decay to be perturbative, which is not applicable during the early stages of reheating, but is necessary to completely drain out the inflaton energy density. A detailed study of such non-perturbative effects require dedicated lattice simulations, that is beyond the scope of this work.

The paper is organized as follows. In Sec.~\ref{sec:model} describe the model of leptogenesis. The the generation of baryon asymmetry and DM abundance during reheating is discussed in Sec.~\ref{sec:cogenesis}. Testability of the model with inflationary GW is explained in Sec.~\ref{sec:pgw}. Finally, we conclude with Sec.~\ref{sec:concl}. 
\section{The set-up}
\label{sec:model}
We study the vanilla leptogenesis scenario for which we extend the SM particle content with the addition of two generations of SM gauge singlet RHNs denoted by $N_R$. The corresponding Lagrangian is given by
\begin{align}\label{eq:yukawa}
&\mathcal{L}\supset i\,\overline{N_R^c}\,\slashed{\partial}N_R-\Bigl(\frac{1}{2}M_N\overline{N_R^c}N_R+\text{h.c}\Bigr)+\Bigl(-y_N\,\overline{\ell_L}\,\widetilde{\mathbb{H}}\,N_R+\text{h.c.}\Bigr)\,,   
\end{align}
where RHNs are assumed to be mass diagonal and all the generational indices are suppressed. The first term simply corresponds to the kinetic term for singlet fermion. The second term is the lepton number violating Majorana mass term for RHN. The last piece corresponding to the interaction Lagrangian is important since it is responsible for the generation of active neutrino mass, as well as the lepton asymmetry via CP-violating decay of the RHNs, which eventually gets converted into the observed baryon asymmetry via sphalerons. Here $\mathbb{H}$ is the SM Higgs doublet with $\widetilde{ \mathbb{H}}=i\,\sigma_2\,\mathbb{H}^\star$, where $\sigma_2$ is the Pauli spin matrix. Once the neutral component of $\mathbb{H}$ acquires a non-zero vacuum expectation value (VEV) leading to the spontaneous breaking of the $SU(2)_L\otimes U(1)_Y$ gauge symmetry, one can write a Dirac mass term for neutrinos as
\begin{align}
m_D= \frac{y_{N}}{\sqrt{2}}v\,,
\end{align}
where $v=246$ GeV is the SM Higgs VEV. The presence of this Dirac mass, together with the bare Majorana mass for the RHNs, can explain the origin of non-zero light neutrino masses through Type-I seesaw~\cite{Gell-Mann:1979vob, Mohapatra:1979ia},
\begin{align}
m_{\nu}\simeq -m_D\,M_N^{-1}\,m_D^T\,,
\label{NM}
\end{align}
where $m_D \ll M_N$ limit is assumed. The mass eigenvalues are then obtained by diagonalising the light neutrino mass matrix as
\begin{align}
m_{\nu}= \mathcal{U}^*\,m_{\nu}^d\,\mathcal{U}^{\dagger}\,,
\end{align}
with $m_{\nu}^d={\rm diag}\,\left(m_1,\,m_2,\,m_3\right)$ consisting of mass eigenvalues and $\mathcal{U}$ being the Pontecorvo-Maki-Nakagawa-Sakata (PMNS) matrix~\cite{ParticleDataGroup:2022pth} (the charged lepton mass matrix is considered to be diagonal). 

In order to obtain a complex structure of the Yukawa coupling $y_N$ which is essential for CP-violating decay of the RHNs, we adopt the Casas-Ibarra (CI) parametrization~\cite{Casas:2001sr} that allows us to write the Yukawa coupling as,
\begin{align}
y_N = \frac{\sqrt{2}}{v}\,\mathcal{U}\,\sqrt{m_{\nu}^d}\,\mathbb{R}^T\,\sqrt{M_N}
\label{CI}
\end{align}
where $\mathbb{R}$ is a complex orthogonal matrix $\mathbb{R}^T \mathbb{R} = I$, which we choose as
\begin{align}
\mathbb{R} =
\begin{pmatrix}
0 & \cos{z} & \sin{z}\\
0 & -\sin{z} & \cos{z}
\end{pmatrix}\,,
\label{eq:rot-mat}
\end{align} 
where $z=a+ib$ is a complex angle. The diagonal mass matrix $m_{\nu}^d$ is calculable using the latest neutrino oscillation data~\cite{ParticleDataGroup:2020ssz,ParticleDataGroup:2022pth}. Note the presence of two RHN in the present setup compels us to consider the lightest active neutrino mass to be zero. The complex angle $z$ can be chosen in a way that enhances the CP asymmetry while keeping the Yukawa couplings within perturbative limits. The CP asymmetry parameter is defined as
\begin{align}
& \epsilon_i = \frac{\sum_j\Gamma\left(N_i\to\ell_j H\right)-\sum_j\Gamma\left(N_i\to\overline{\ell_j}~ \overline{H}\right)}{\sum_j\Gamma\left(N_i\to\ell_j H\right)+\sum_j\Gamma\left(N_i\to\overline{\ell_j}~ \overline{H}\right)}      
\end{align}
Neglecting flavour effects we can write the final expression for the CP-asymmetry as~\cite{Buchmuller:2004nz,Davidson:2008bu}
\begin{align}\label{eq:cpasym}
	\epsilon_1\simeq\frac{3}{16\pi} \frac{\text{Im}[(y_N^\dagger y_N)^2_{12}]}{(y_N^\dagger y_N)_{11}}\left(\frac{M_{N_1}}{M_{N_2}}\right)\,,
\end{align}
which occurs due to the decay of the lightest right handed neutrino $N_1$. We assume hierarchical Majorana masses for the RHNs: $M_{N_1}\ll M_{N_{2}}$. Here onward we will denote the lightest neutrino mass as $M_{N_1}\equiv M_N$. Finally,  we consider lepton-number-violating interactions of $N_1$ rapid enough to wash out the lepton number asymmetry originated by the $N_2$. Therefore, only the CP-violating asymmetry from the decay of $N_1$ survives and is relevant for leptogenesis. In the following analysis we will consider the $N_1$ is produced during the period of reheating entirely from the thermal bath, and further decays to produce the asymmetry via vanilla leptogenesis.   
\subsection{Inflaton dynamics during reheating}
\label{sec:reheat}
At the end of inflation, oscillation of the inflaton occurs at the bottom of a potential $V(\phi)$ of the form
\begin{equation}\label{eq:inf-pot}
    V(\phi) = \lambda\, \frac{\phi^n}{\Lambda^{n - 4}}\,,
\end{equation}
where $\lambda$ is a dimensionless coupling, $\Lambda$ is the scale of inflation and $\phi$ is the inflaton field. The nature of the inflaton potential during reheating, as it is understandable, is decided by $n>0$. Such a potential naturally arise in a number of inflationary scenarios, for example, the $\alpha$-attractor T- or E-models~\cite{Kallosh:2013hoa, Kallosh:2013yoa, Kallosh:2013maa}, or the Starobinsky model~\cite{Starobinsky:1980te, Starobinsky:1981vz, Starobinsky:1983zz, Kofman:1985aw}. Now, given a particular inflationary model, for example, in $\alpha$-attractor T-model~\cite{Kallosh:2013hoa, Kallosh:2013yoa}
\begin{equation}
    V(\phi ) =\lambda\, M_P^4 \left[\tanh \left(\frac{\phi}{\sqrt{6\, \alpha}\, M_P}\right)\right]^n \simeq \lambda\, M_P^4 \times
    \begin{cases}
        1 & \; \text{for}\; \phi \gg M_P,\\
        \left(\frac{\phi}{\sqrt{6\,\alpha}\,M_P}\right)^n & \; \text{for}\; \phi\ll M_P\,,
    \end{cases}
\end{equation}
where $M_P$ is the reduced Planck mass. The overall scale of the potential parameterized by the coupling $\lambda\simeq 18\pi^2 A_S/\left(6^{n/2}\,N_\star^2\right)$ is determined from the amplitude of the CMB power spectrum $\ln(10^{10}A_S)=3.044$~\cite{Planck:2018jri} and the number of e-folds measured from the end of inflation to the time when the pivot scale $k_*=0.05~{\rm Mpc}^{-1}$ exits the horizon $N_\star=55$. 
The equation of motion for the oscillating inflaton field reads~\cite{Turner:1983he}
\begin{equation} \label{eq:eom0}
    \ddot\phi + (3\, H + \Gp)\, \dot\phi + V'(\phi) = 0\,,
\end{equation} 
where dots represent derivative with respect to time, while the prime is the derivative with respect to the field. Here $H$ is the Hubble parameter, $\Gp$ the rate at which inflaton energy density is transferred into the energy density of its decay products. We will elaborate more on this in a moment. 

Defining the energy density and pressure of $\phi$ as $\rp \equiv \frac12\, \dot\phi^2+ V(\phi)$ and $p_\phi \equiv \frac12\, \dot\phi^2 - V(\phi)$, together with the equation of state (EoS) parameter $w \equiv p_\phi/\rp = (n - 2) / (n + 2)$~\cite{Turner:1983he}, one can write the evolution of the inflaton energy density as
\begin{equation} \label{eq:drhodt}
    \dot\rp + \frac{6\, n}{2 + n}\, H\, \rp = - \frac{2\, n}{2 + n}\, \Gp\, \rp\,,
\end{equation}
where the ``dots" represent derivative with respect to time $t$. During reheating $a_I \ll a \ll \arh$, where $a$ is the scale factor, the term associated with expansion, $H\, \rp$ typically dominates over the interaction term $\Gp\, \rp$. Then it is possible to solve Eq.~\eqref{eq:drhodt} analytically, leading to
\begin{equation} \label{eq:rpsol}
    \rp(a) \simeq \rp (\arh) \left(\frac{\arh}{a}\right)^\frac{6\, n}{2 + n}.
\end{equation}
Here, $a_I$ and $\arh$ correspond to the scale factor at the end of inflation and at the end of reheating, respectively. Note that, for $n=2$, the inflaton energy density scales as $a^{-3}$, same as that of a pressureless matter. Since the Hubble rate during reheating is dominated by the inflaton energy density, it follows that
\begin{equation} \label{eq:Hubble}
    H(a) \simeq H(\arh) \times
    \begin{cases}
        \left(\frac{\arh}{a}\right)^\frac{3\, n}{n + 2} &\text{ for } a \leq \arh\,,\\
        \left(\frac{\arh}{a}\right)^2 &\text{ for } \arh \leq a\,.
    \end{cases}
\end{equation}
At the end of the reheating ($a = \arh$), the energy densities of the inflaton and radiation are equal, $\rR(\arh) = \rp(\arh) = 3\, M_P^2\, H(\arh)^2$. The evolution of the SM radiation energy density $\rR$, on the other hand, is governed by the Boltzmann equation of the form~\cite{Garcia:2020wiy}
\begin{equation} \label{eq:rR}
    \dot\rR + 4\, H\, \rR =  \frac{2\, n}{2 + n}\, \Gp\, \rp\,.
\end{equation}
The effective mass $m_\phi(a)$ for the inflaton can be obtained from the second derivative of Eq.~\eqref{eq:inf-pot}, which reads
\begin{align}\label{eq:inf-mass1}
& m_\phi(a)^2 \equiv \frac{d^2V}{d\phi^2} = n\, (n - 1)\, \lambda\, \frac{\phi^{n - 2}}{\Lambda^{n - 4}}\simeq n\, (n-1)\, \lambda^\frac{2}{n}\, \Lambda^\frac{2\, (4 - n)}{n} \rp(a)^{\frac{n-2}{n}}\,,
\end{align}
or, equivalently,
\begin{align}\label{eq:inf-mass2}
m_\phi(a) \simeq m_I \left(\frac{a_I}{a}\right)^\frac{3 (n-2)}{n+2}\,,    
\end{align}
where $m_I\equiv m_\phi(a_I)\simeq 1.5\times10^{13}$ GeV. Note, for $n=2$, the inflaton mass is constant $m_\phi\simeq m_I$, while for $n>2$, $m_\phi$ has a field dependence, as a consequence of which the mass decreases with time (scale factor). 

In order to solve Eq.~\eqref{eq:drhodt} and \eqref{eq:rR}, we need the details of the inflaton dissipation rate. This is governed by the underlying particle physics model involving the inflaton and the SM fields. Here we consider two scenarios:
\begin{itemize}
\item In the first case, the inflaton decays into a pair of SM-like fermions $\Psi$, which are part of the radiation bath\footnote{
 Gauge-invariant interaction between the inflaton and the SM fermions can be realized in a more realistic set-up. For example, the inflaton can be promoted to an electroweak multiplet~\cite{Chen:2010uc, Borah:2018rca} or electroweak gauge symmetry may not be restored at high temperatures \cite{Meade:2018saz}.  We, however, remain agnostic about the microscopic model and write an effective inflaton-SM interaction as in Refs.~\cite{Kaneta:2019zgw,Garcia:2020wiy,Datta:2023pav}.}, via the Yukawa interaction 
\begin{align}
\mathcal{L}\supset - y_\psi\, \overline{\Psi}\, \Psi\, \phi\,, 
\end{align}
with a decay rate
\begin{equation} \label{eq:fer_gamma}
    \Gp(a) = \frac{\yeff^2}{8\pi}\, m_\phi(a)\,,
\end{equation}
where the effective coupling $\yeff \ne y_\psi$ (for $n \neq 2$) is obtained after averaging over several oscillations~\cite{Shtanov:1994ce, Ichikawa:2008ne, Garcia:2020wiy} (see Appendix~\ref{sec:inf-decay} for details). Here we assume the SM bath thermalizes instantaneously. The evolution of the SM energy density in this case becomes~\cite{Bernal:2022wck}
\begin{equation} \label{eq:rR_fer}
    \rR(a) \simeq \frac{3\, n}{7 - n}\, M_P^2\, \Gp(\arh)\, H(\arh) \left(\frac{\arh}{a}\right)^\frac{6 (n - 1)}{n+2} \left[1 - \left(\frac{a_I}{a}\right)^\frac{2 (7 - n)}{2 + n}\right]\,,
\end{equation}
implying the evolution of corresponding bath temperature $T(a)\simeq\Trh\, \left(\arh/a\right)^{\frac{3\,(n-1)}{2\,(n+2)}}$.
\item For the second case, we consider the inflaton to decay into a pair of bosons through the trilinear scalar interaction 
\begin{equation}\label{eq:int1}
    \mathcal{L}\supset -\mu\, \phi\, |\varphi|^2\,,  
\end{equation}
where $\varphi$ can be identified as the SM-like Higgs. In this case, the decay rate becomes
\begin{equation} \label{eq:bos_gamma}
       \Gp(a) = \frac{\mueff^2}{8\pi\, m_\phi(a)}\,,
\end{equation}
where again the effective coupling $\mueff \ne \mu$ (if $n\neq2$) can be obtained after averaging over oscillations\footnote{We have assumed the thermal masses of the daughter particles to be negligible, which remains a valid assumption as long as $T < m_\phi/g $, with $g$ being the typical SM gauge coupling strength~\cite{Kolb:2003ke}. This however is true when the inflaton potential is quadratic during reheating. 
A detailed analysis of thermal mass corrections for a general monomial potential is beyond the scope of the present analysis.}. The SM energy density reads
\begin{equation} \label{eq:rR_bos}
    \rR(a) \simeq \frac{3\, n}{1 + 2\, n}\, M_P^2\, \Gp(\arh)\, H(\arh) \left(\frac{\arh}{a}\right)^\frac{6}{2 + n} \left[1 - \left(\frac{a_I}{a}\right)^\frac{2\, (1 + 2 n)}{2 + n}\right]\,,
\end{equation}
with which the SM bath temperature during reheating evolves as: $T\simeq\Trh\,\left(\arh/a\right)^{\frac{3}{2\,(n+2)}}$. In the following analysis, we will consider, each reheating scenario to be present at a time. The bath temperature during reheating therefore has a general form:
\begin{align}\label{eq:Tevol}
& T(a)\simeq\Trh\,\left(\frac{\arh}{a}\right)^\xi\,,    
\end{align}
where $\xi=3\,(n-1)/(2\,(n+2))$ for fermionic reheating, while for bosonic reheating $\xi=3/(2\,(n+2))$. Note that, for the bosonic case, corresponding inflaton coupling has a mass dimension. 
\end{itemize}
To avoid affecting the BBN predictions, the reheating temperature $\Trh$ must satisfy $\Trh > T_\text{BBN} \simeq 4$~MeV~\cite{Sarkar:1995dd, Kawasaki:2000en,Hannestad:2004px, DeBernardis:2008zz, deSalas:2015glj,Hasegawa:2019jsa}.
Finally, we introduce the following Yukawa-like interaction between the inflaton and the RHNs
\begin{align}
\mathcal{L}_{\phi NN}\supset -y_{\phi N}\,\phi\,\overline{N^c}\,N+\text{h.c.}\,,
\end{align}
which is allowed since all the fields involved are singlets under the SM gauge symmetry and there is no underlying symmetry to forbid this interaction. The coupling $y_{\phi N}$ is a free parameter that we will constrain from the requirement of successful reheating and from the generation of right baryon asymmetry. This interaction results in production of RHNs from on-shell 2-body decay of the inflaton as long as $m_\phi(a)>2\,M_N$. However, for $n>2$, there exists a particular scale factor, given by
\begin{align}\label{eq:ast}
a_\star= a_I\,\left(\frac{M_N}{2\,m_I}\right)^\frac{n+2}{3\,(2-n)}\,,    
\end{align}
above which $m_\phi(a)<M_N/2$, and the $\phi\to NN$ production channel is kinematically forbidden. Note, the above relation is not valid for $n=2$ since the inflaton mass is independent of the scale factor in that case. This interaction opens up the following $2\to2$ production channels: (i) SM\,SM $\to NN$ ($s$-channel inflaton mediation) and (ii) $\phi\phi\to NN$ ($t$-channel RHN mediation). Note that, the former one is always kinematically feasible as long as the center of mass energy is sufficiently large to pair-produce the RHNs, i.e., $s>4\,M_N^2$. Whereas, for the later, one needs to have $m_\phi>M_N$.

\begin{table}[htb!]
    \begin{center}
        \begin{tabular}{|c|c|c|c|}
            \hline
            $T(a)$ & $n=2$ & $n=4$ & $n=6$  \\
            \hline\hline
            Fermionic & $a^{-3/8}$ & $a^{-3/4}$ & $a^{-15/16}$ \\ 
            \hline
            Bosonic & $a^{-3/8}$ & $a^{-1/4}$ & $a^{-3/16}$ \\
            \hline
        \end{tabular}
    \end{center}
    \caption {Temperature evolution of radiation bath during reheating for different choices of $n$.}
    \label{tab:T}
\end{table}

In Tab.~\ref{tab:T} we have listed the scale-factor dependence of the bath temperature during reheating for different choices of $n$, corresponding to bosonic and fermionic reheating. For $n=2$, the bath evolves identically in either reheating scenarios. For $n>4$, in case of fermionic reheating, the temperature falls approximately inverse to the scale factor, i.e., $T\propto a^{-1}$, which is the characteristic of radiation dominated background.
\begin{figure}[htb!]
    \centering
    \includegraphics[scale=0.37]{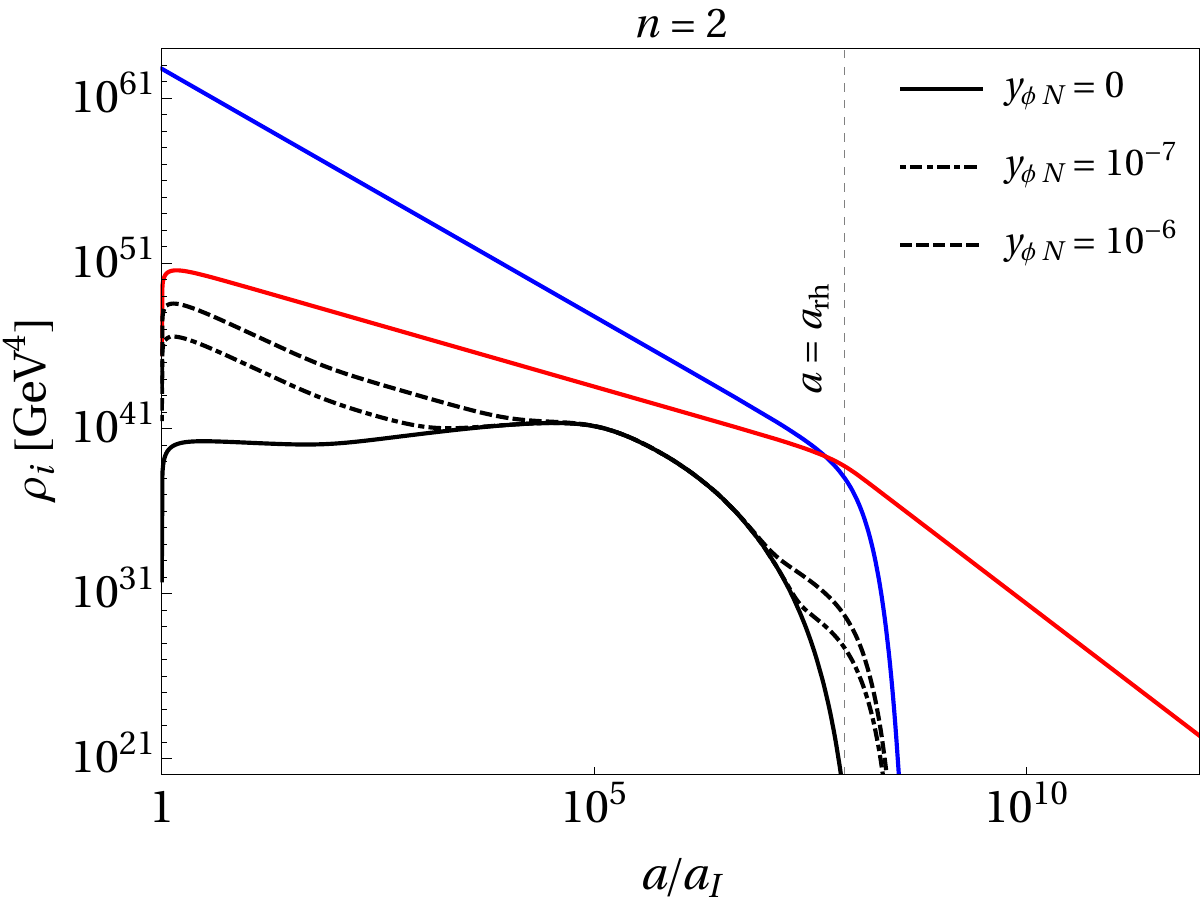}~\includegraphics[scale=0.37]{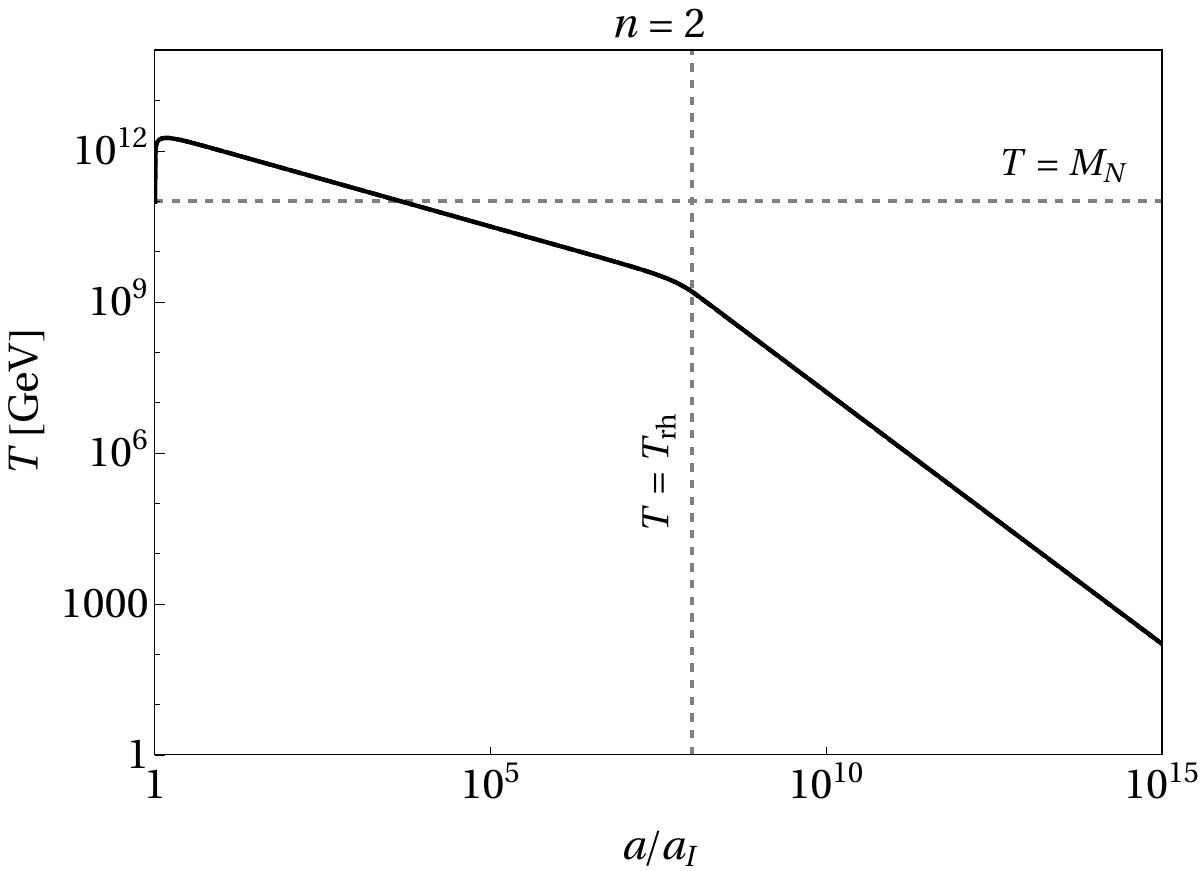}\\[10pt]
    \includegraphics[scale=0.37]{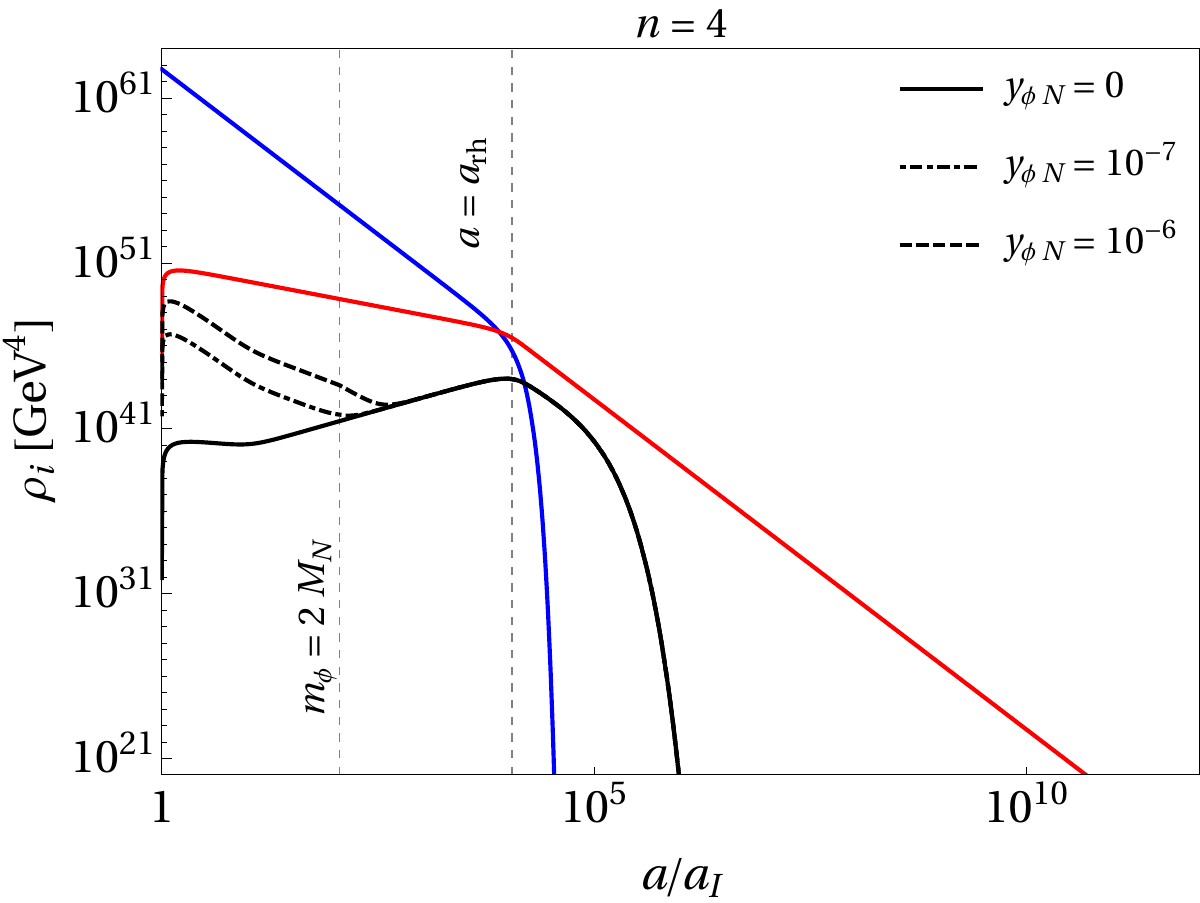}~\includegraphics[scale=0.37]{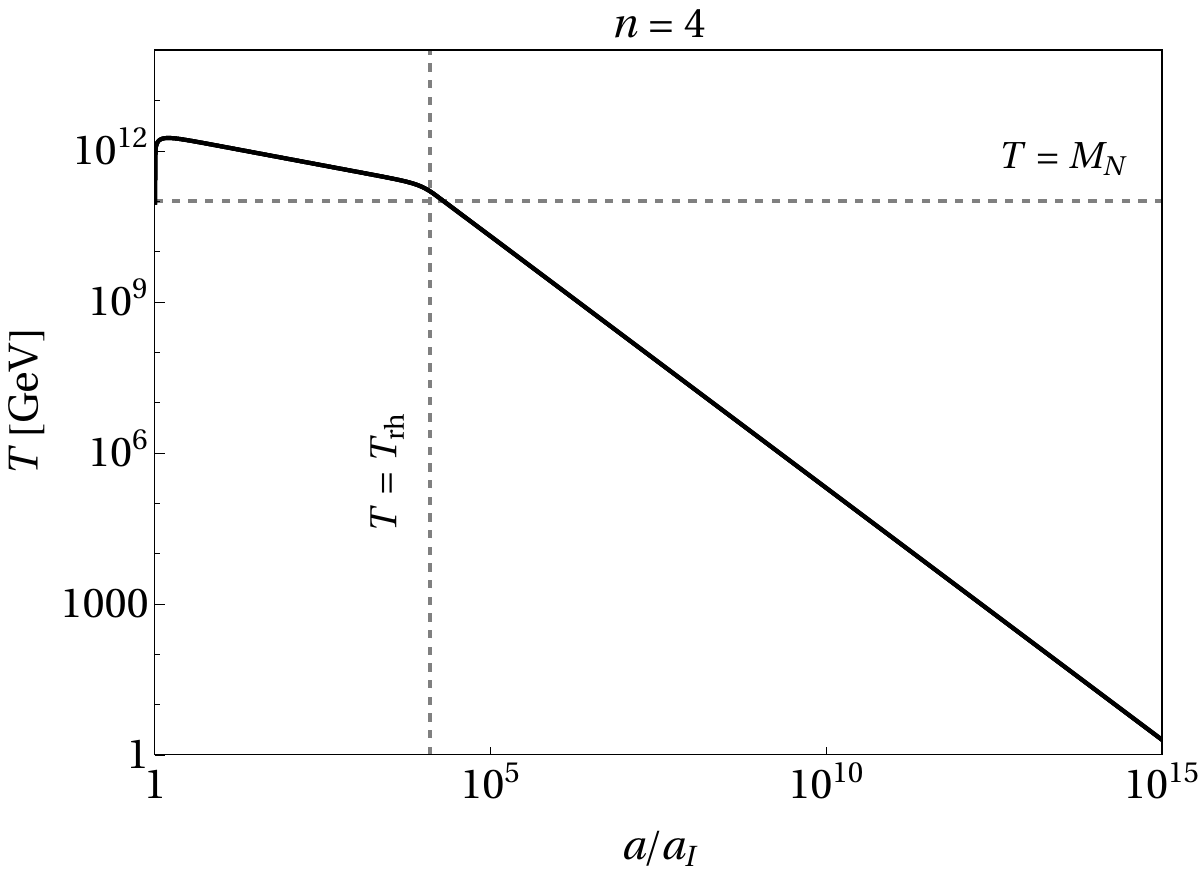}
    \\[10pt]
    \includegraphics[scale=0.37]{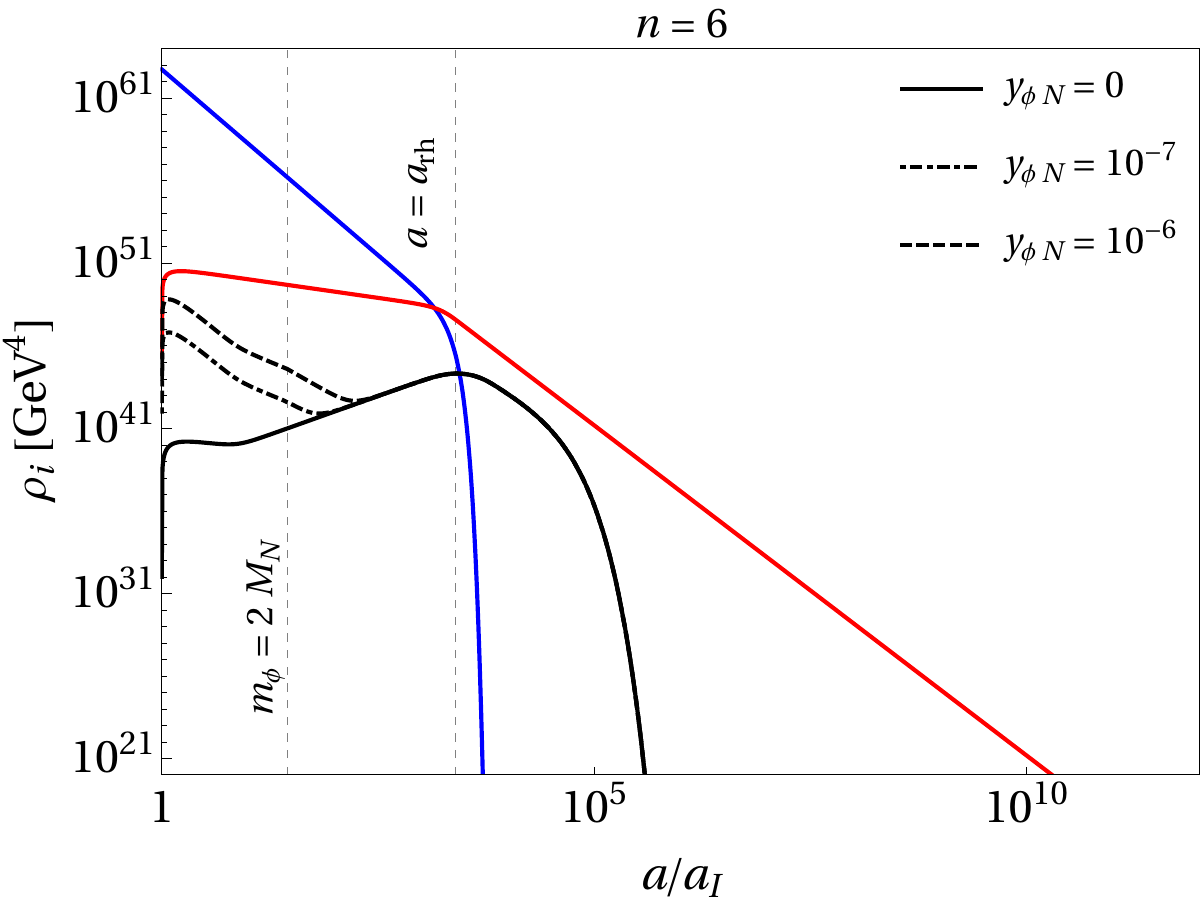}~\includegraphics[scale=0.37]{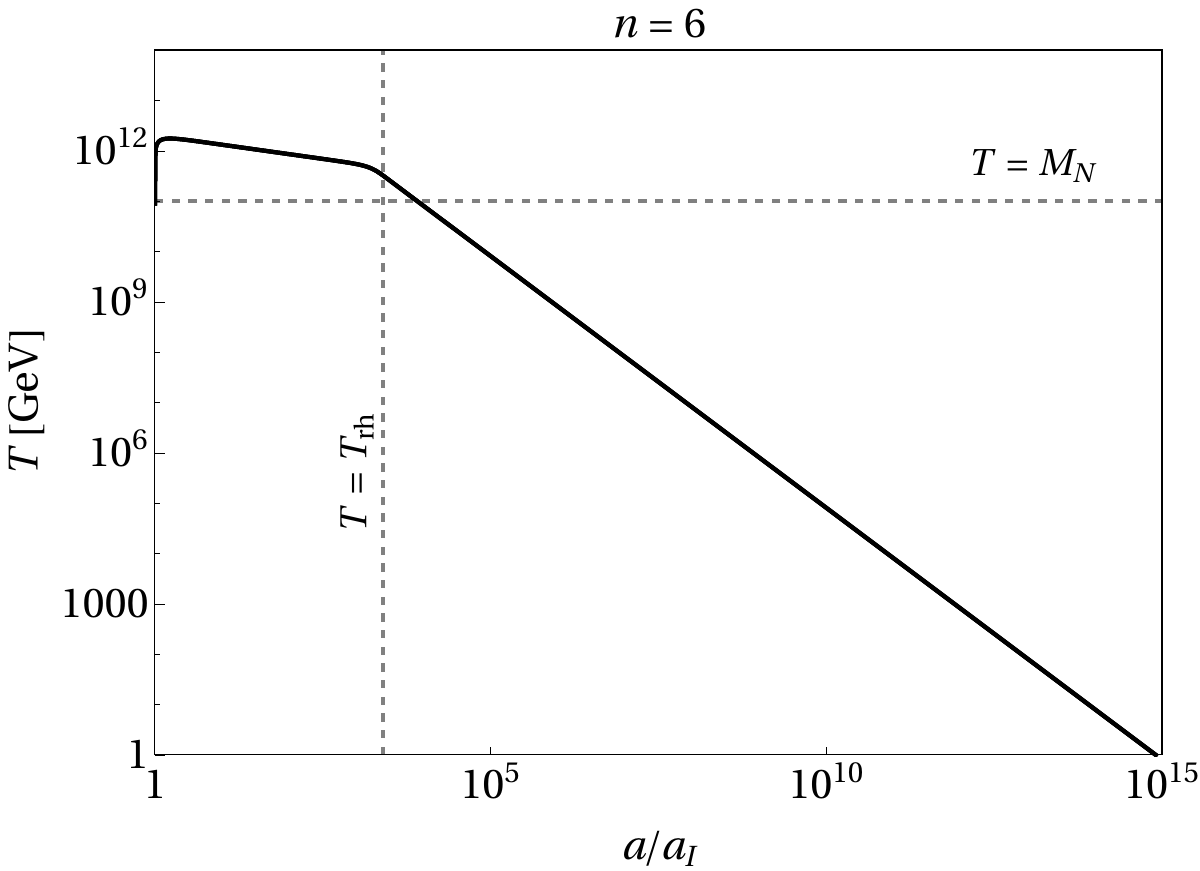}
    \caption{Bosonic reheating scenario. Left column: Evolution of energy densities of inflaton (blue solid), RHN (blue dot-dashed) and radiation (red) with the scale factor for different choices of $n$. The solid, dot-dashed, dashed and dotted curves correspond to $\yNN=\{0,\,10^{-7},\,10^{-6}\}$, respectively. Right column: Corresponding evolution of SM bath temperature in each case.  Here we fix $\meff=10^{8}$ GeV and $M_N=10^{11}$ GeV.}
    \label{fig:rho-T1-b}
\end{figure}
\begin{figure}[htb!]
    \centering
    \includegraphics[scale=0.37]{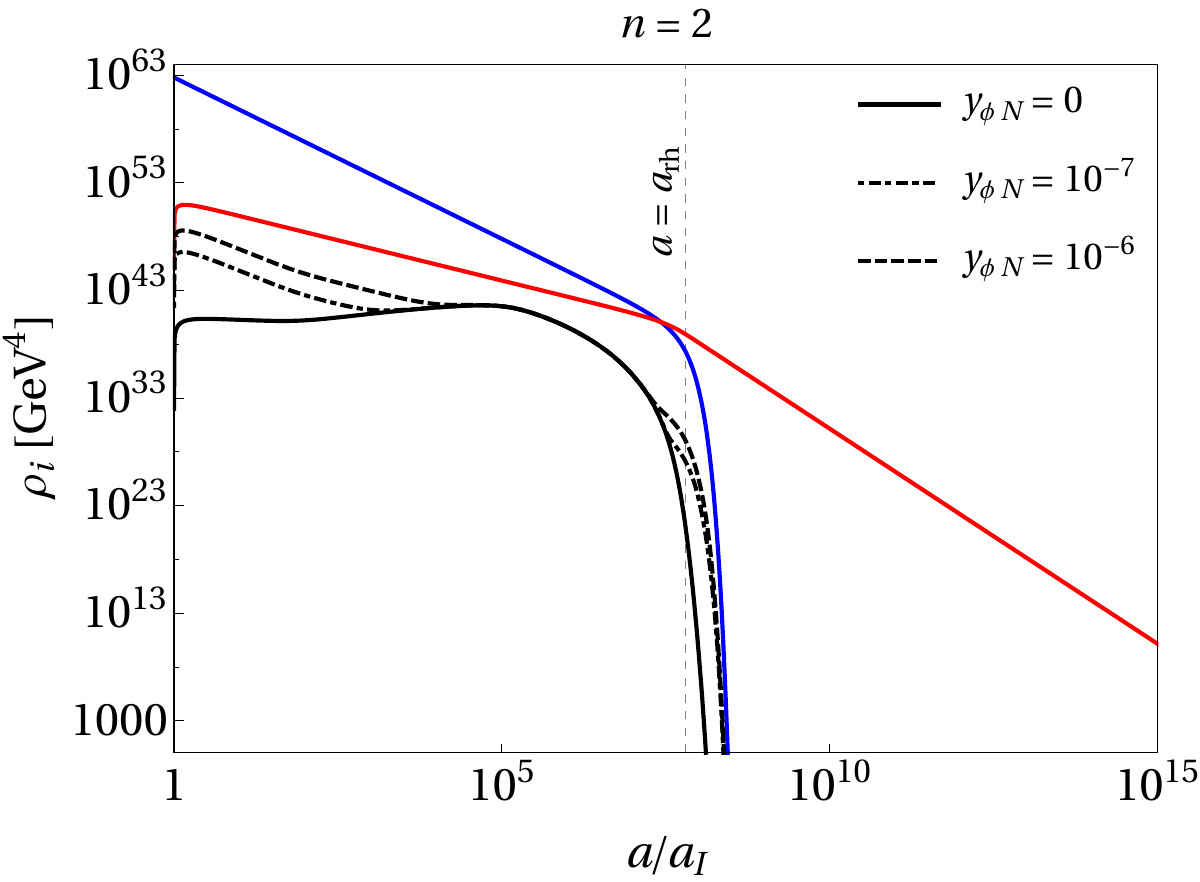}~\includegraphics[scale=0.37]{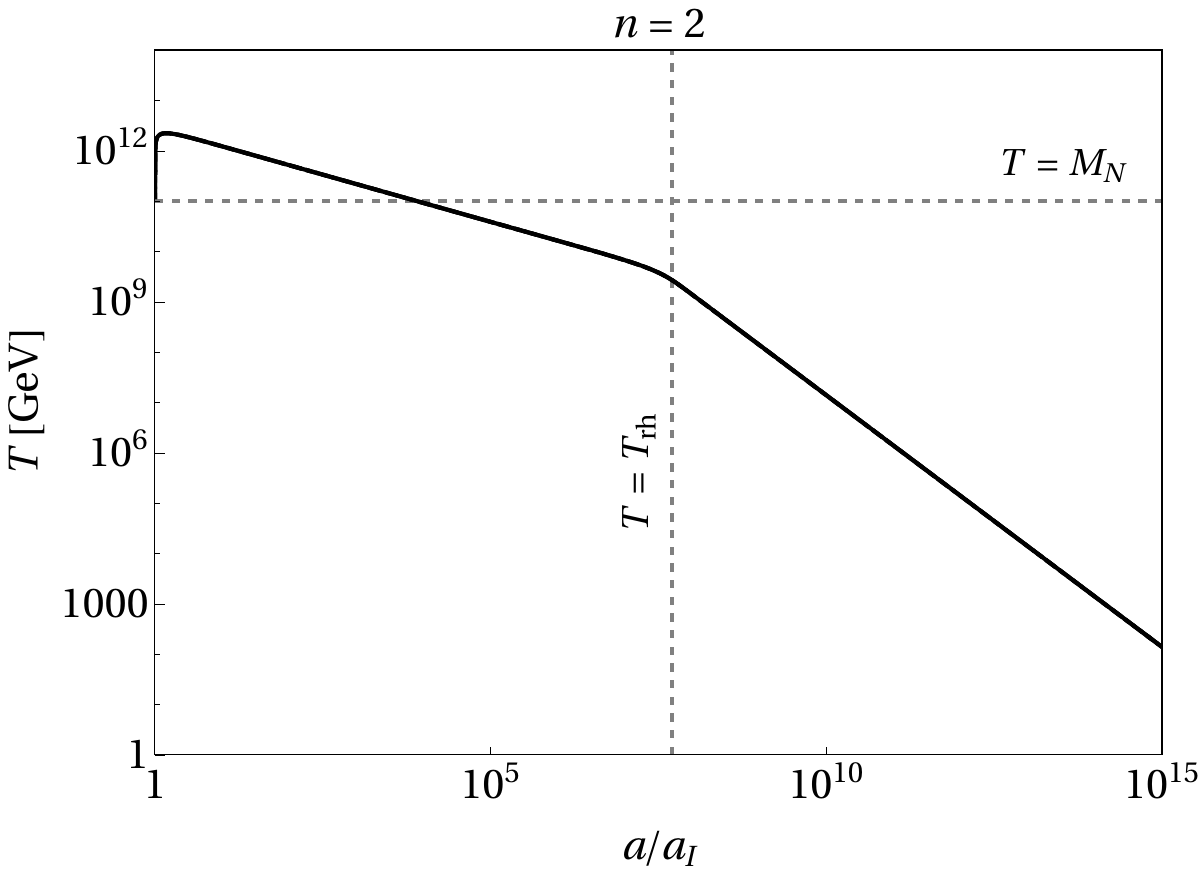}\\[10pt]
    \includegraphics[scale=0.37]{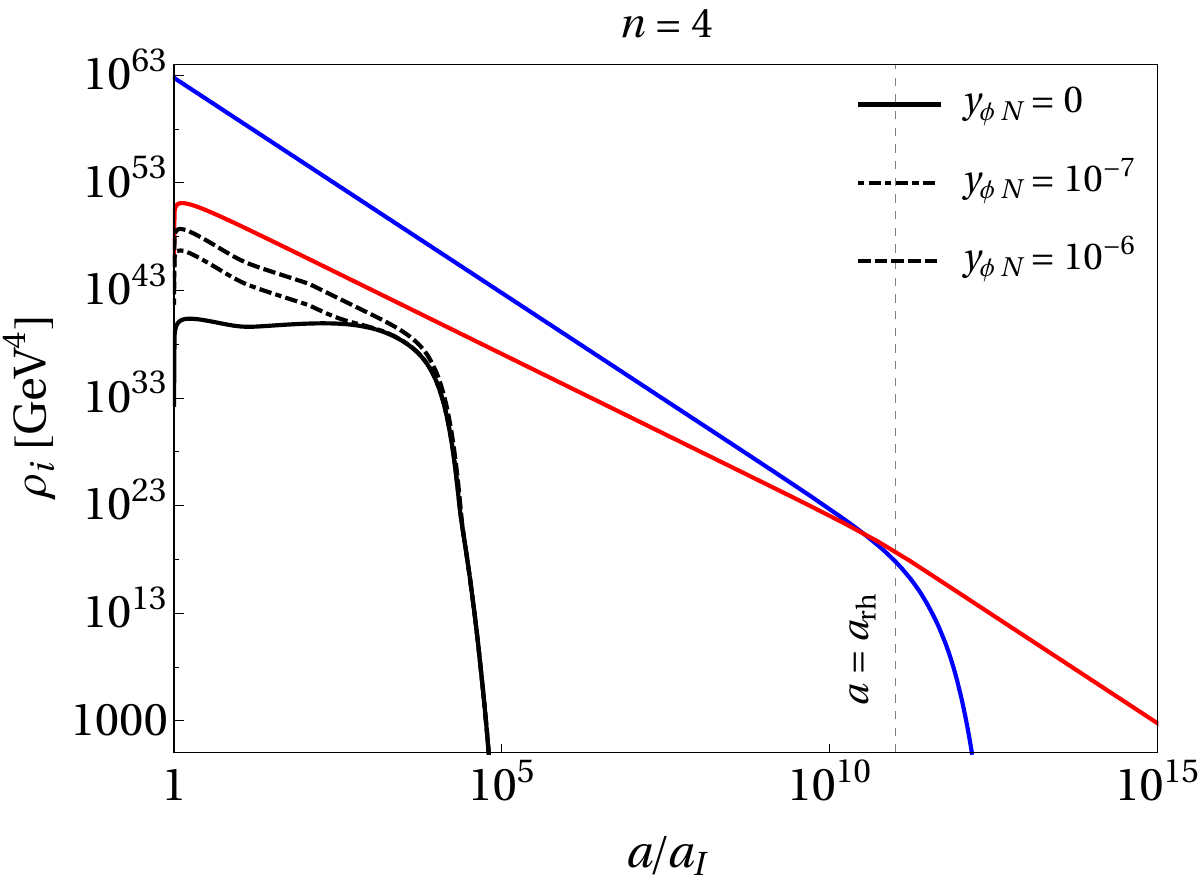}~\includegraphics[scale=0.37]{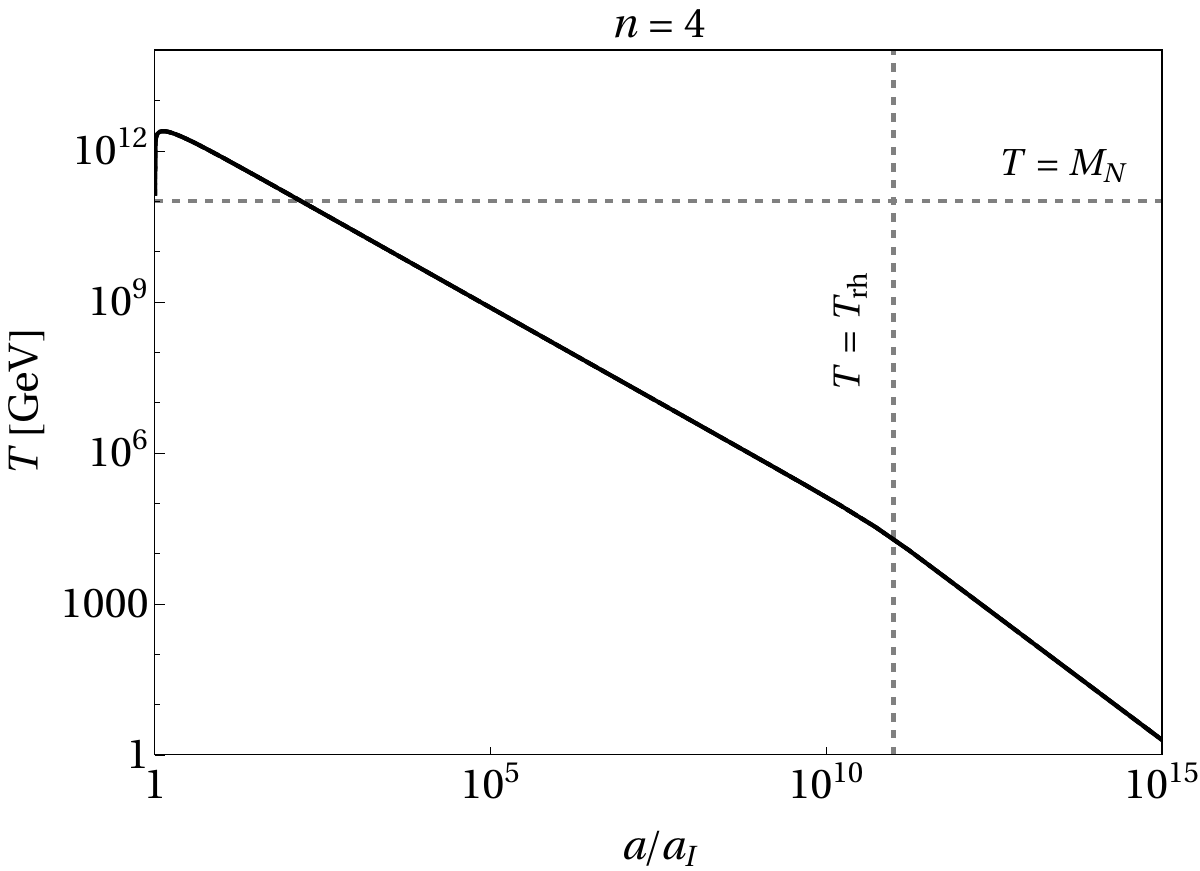}
    \\[10pt]
    \includegraphics[scale=0.37]{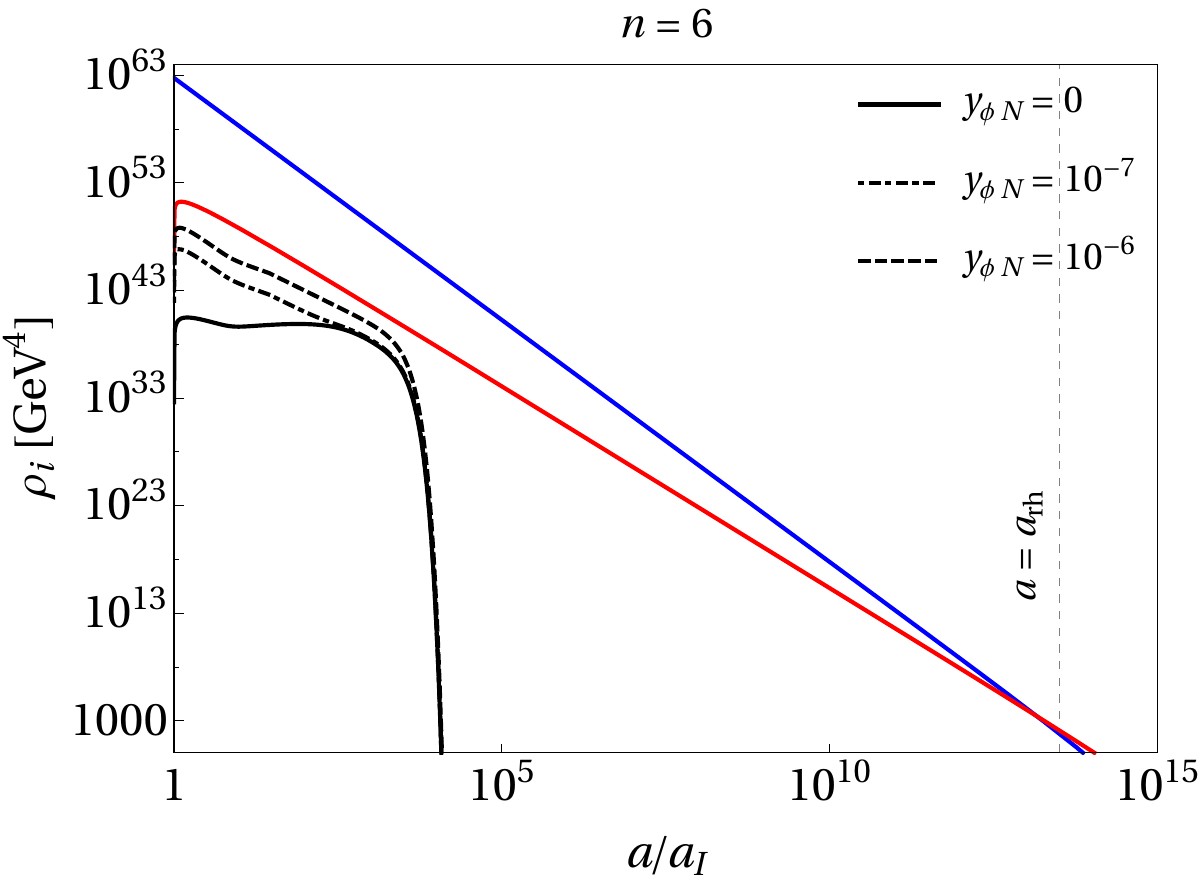}~\includegraphics[scale=0.37]{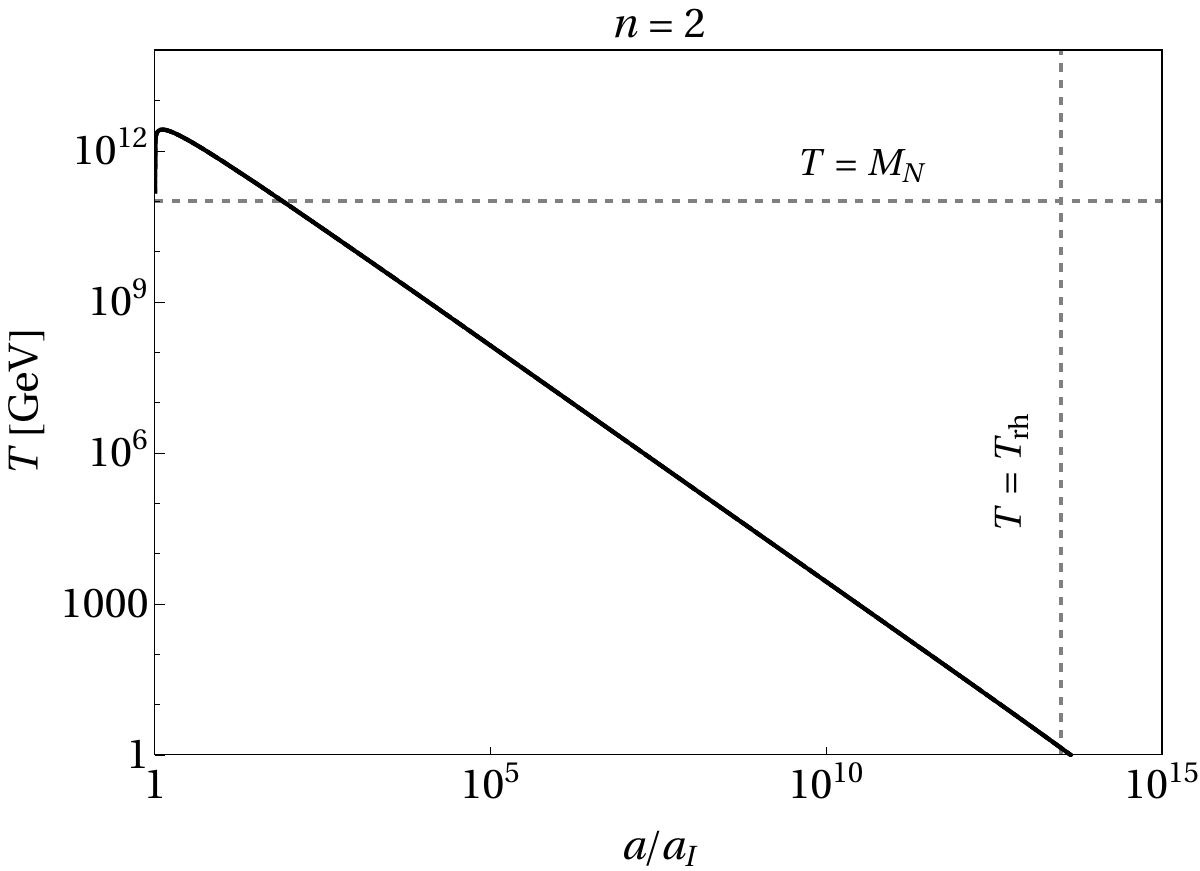}
    \caption{Fermionic reheating scenario. Left column: Evolution of energy densities of inflaton (blue solid), RHN (blue dot-dashed) and radiation (red) with the scale factor for different choices of $n$. The solid, dot-dashed, dashed and dotted curves correspond to $\yNN=\{0,\,10^{-7},\,10^{-6}\}$, respectively. Right column: Corresponding evolution of SM bath temperature in each case. Here we fix $\yeff=10^{-5}$ and $M_N=10^{11}$ GeV.}
    \label{fig:rho-T1-f}
\end{figure}
In the left panels of Fig.~\ref{fig:rho-T1-b} and \ref{fig:rho-T1-f}, we show the evolution of different components, namely, inflaton energy density, radiation energy density, and RHN energy density as a function of the scale factor, for bosonic and fermionic reheating scenarios, respectively. From the right panel of either figure, one can see that the scaling of the bath temperature $T(a)$ matches exactly with those obtained analytically in Tab.~\ref{tab:T}. As the inflaton decay width for fermionic reheating $\propto m_\phi(a)$, while for bosonic final states it is $\propto 1/m_\phi(a)$, the reheating process becomes more efficient over time for the bosonic final states than for the fermionic final states for $n>2$, since the inflaton mass $m_\phi(a)$ is a decreasing function of time for $n>2$.

For both bosonic and fermionic reheating, we note, $\yNN>0$ results in an increase of $\rho_N$, simply because it adds an additional channel to RHN production. By comparing the inflaton decay widths into SM and into RHN final states, it is possible to obtain a lower bound on inflaton-RHN coupling as,
\begin{align}
& \yNN>
\begin{cases}
\mu/\sqrt{m_\phi\,M_N}\,, & \text{for bosonic reheating}
\\[10pt]
y\,\sqrt{m_\phi/M_N}\,, & \text{for fermionic reheating}\,,
\end{cases}
\end{align}
such that inflaton decays dominantly into RHN final states. Considering bulk of the RHNs are produced from $\phi\to NN$ at the very beginning of reheating  (when $m_\phi=m_I$), we see, $\yNN\gtrsim 10^{-4}$ for $M_N=10^{11}$ GeV. To ensure $\phi\to NN$ decay is always perturbative, we choose $\yNN\lesssim 10^{-5}$~\cite{Drewes:2017fmn,Drewes:2019rxn,Garcia:2020wiy} and hence, the inflaton decays dominantly into the SM producing the radiation bath. For our choice of parameters, irrespective of the choice of $n$, the onset of radiation domination is therefore set by the approximate equality between inflaton decay rate and the Hubble rate $\Gp(\arh)\simeq H(\arh)$, where $\Gp$ corresponds to inflaton decay into SM final states. Clearly, one can then trade the inflaton-SM couplings for $\Trh$, and it is easy to understand that a larger coupling shall result in higher $\Trh$. One can then also obtain an approximate lower bound on the couplings by demanding $\Trh\gtrsim 4$ MeV, that results in $\yeff\gtrsim 10^{-7}$ and $\mueff\gtrsim 10^4$ GeV, for $n=2$. For $n>2$, this bound is rather relaxed for fermionic reheating as $\Gp\propto m_\phi$ and more stringent for bosonic reheating since $\Gp\propto 1/m_\phi$, and $m_\phi$ decreases with the scale factor for $n>2$. It is important to note here, for bosonic reheating scenario, the RHN decay is completed before the end of reheating for $n=2$. For $n=4$ and $n=6$, however, the reheating ends earlier and the asymmetry is produced during radiation domination. In case of fermionic reheating, the asymmetry is always produced during reheating. 

\begin{figure}[htb!]
    \centering
    \includegraphics[scale=0.375]{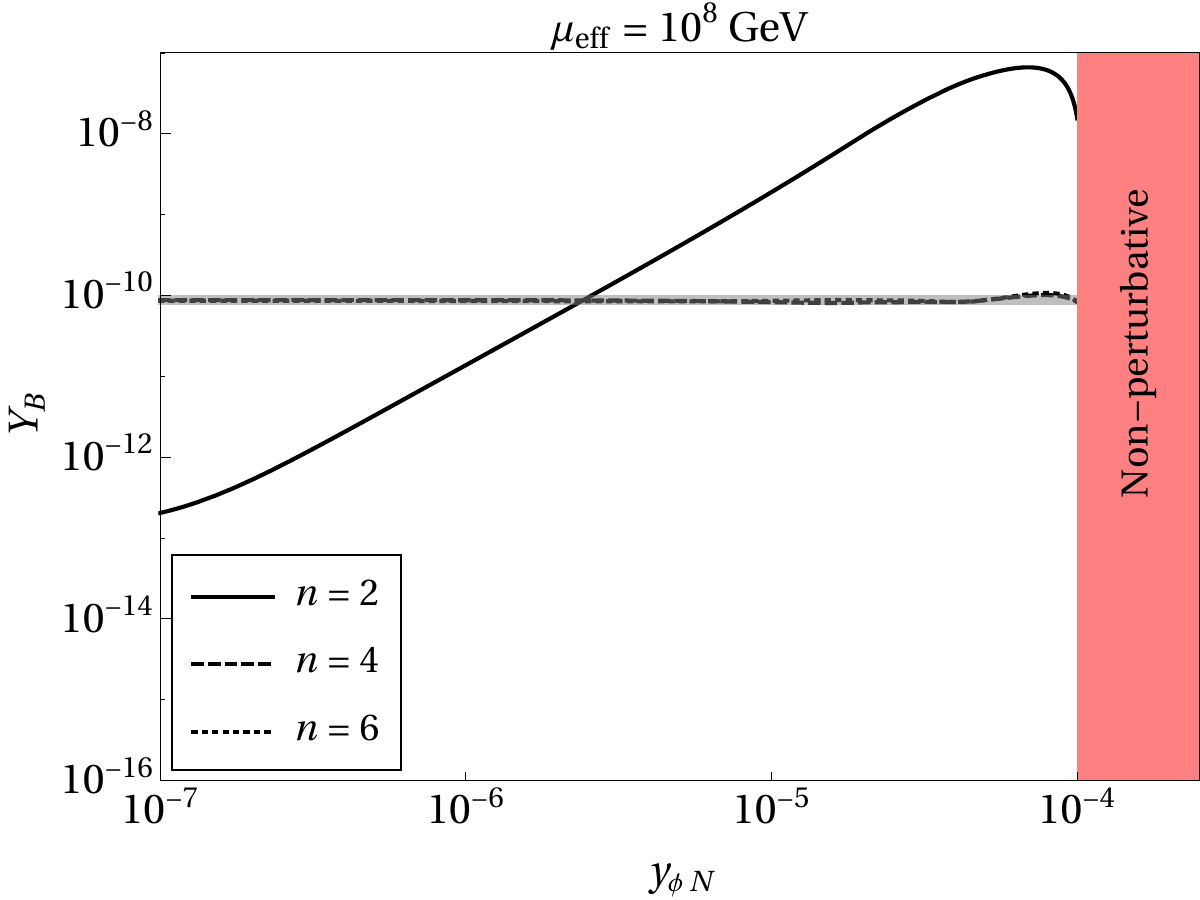}~\includegraphics[scale=0.375]{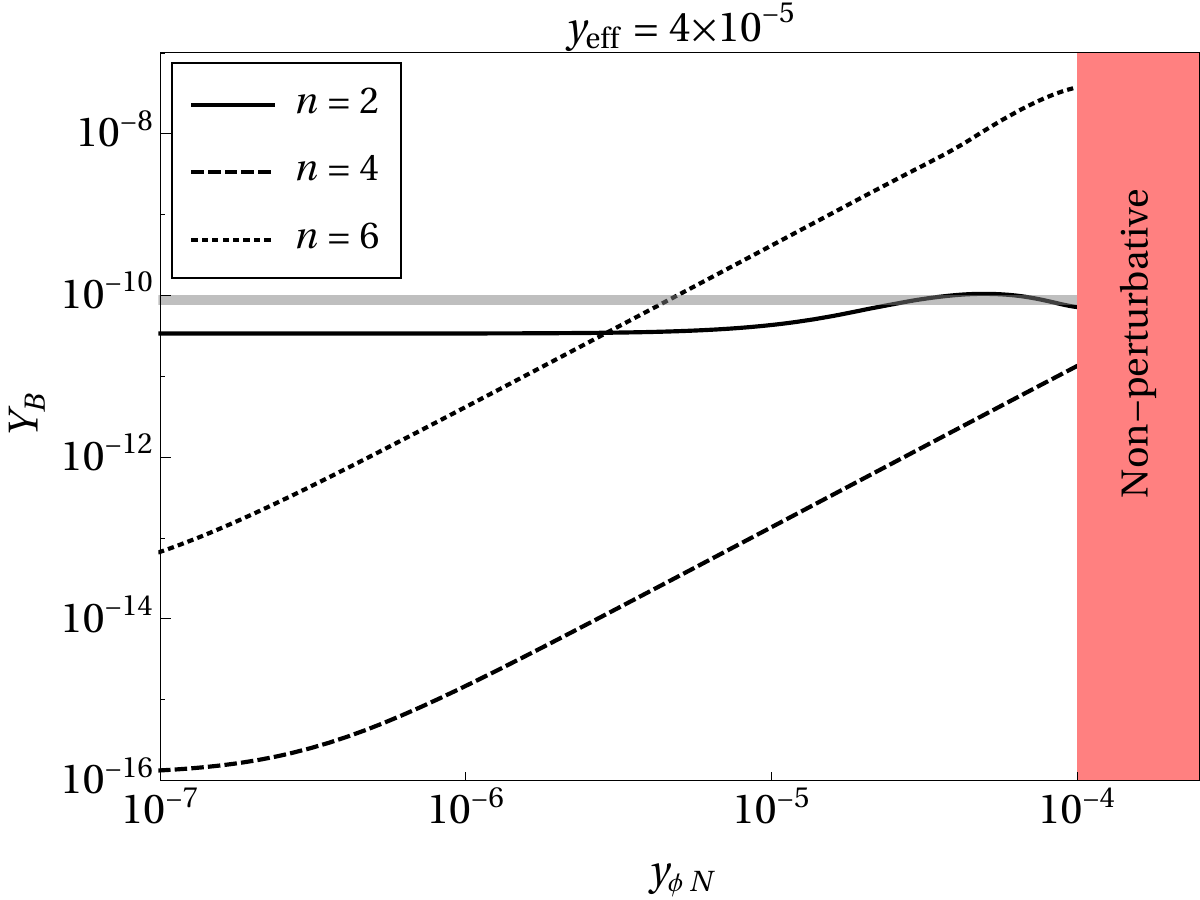}
    \caption{Baryon asymmetry as a function of $\yNN$ for $M_N=10^{11}$ GeV, where in the left panel we consider bosonic and in the right panel fermionic reheating, respectively. Different curves correspond to different choices of $n$, as mentioned in the plot legend. The gray horizontal line corresponds to present day baryon asymmetry. The red shaded region is disallowed from the requirement of perturbative decay of the inflaton (see text for details).}
    \label{fig:asym}
\end{figure}

It is imperative to underscore the necessity of incorporating the dynamics of the preheating phase. It is already established that, during the period of reheating, the equation of state (EoS) parameter asymptotically approaches $w \to 1/3$ for $n \gtrsim 3$, a consequence of inflaton fragmentation and parametric resonance~\cite{Amin:2010dc,Lozanov:2019jxc,Garcia:2023eol,Garcia:2023dyf}. This suggests that preheating plays a crucial role in efficiently transitioning the Universe into a radiation-dominated phase. However, the complete dissipation of the inflaton’s energy necessitates perturbative decay through trilinear couplings between the inflaton and its daughter particles, a process anticipated to dominate the final stages of the post-inflationary heating phase~\cite{Kofman:1997yn,Kofman:1985aw}. Furthermore, as it has been shown in~\cite{Drewes:2017fmn,Drewes:2019rxn} in context with $\phi^2$ potential, perturbative reheating is valid for inflaton-scalar coupling  $\mu/m_\phi\lesssim 10^{-5}$ or inflaton-fermion coupling $y_\psi\lesssim 10^{-5}$, which we have considered for $n=2$ scenario in Fig.~\ref{fig:asym}. However, for $n>2$, the non-perturbative effects can become important. Such analysis is beyond the scope of the present paper.


\section{Leptogenesis and DM production during reheating}
\label{sec:cogenesis}
In this section, we discuss generation of baryon asymmetry during the epoch of reheating via leptogenesis. We consider production of RHNs from the thermal bath during reheating via the inverse decay process $\ell\,H\to N$, and their subsequent decay results in the generation of baryon asymmetry. In order to track the evolution of individual components, we solve the following set of coupled BEQs involving inflaton energy density $\rp$, radiation energy density $\rho_R$, RHN number density $n_{N_1}$, $B-L$ number density $n_{B-L}$ generated from the RHN decay
\begin{align}\label{eq:BEQ-phiNN}
& \dot\rho_\phi+3\,H\,(1+w)\,\rho_\phi=-\left(\Gamma_{\phi\to ii}+\Gamma_{\phi\to N_1N_1}\right)\,(1+w)\,\rho_\phi   
\nonumber\\&
\dot\rho_R + 4\,H\,\rho_R = \Gamma_{\phi\to ii}\,(1+w)\,\rho_\phi+\langle \Gamma_N\rangle\,\left(n_{N_1}-n_{N_1}^{\rm eq}\right)
\nonumber\\&
\Dot{n}_{N_1}+ 3\,H\,n_{N_1}= - \langle \Gamma_N\rangle\,\left(n_{N_1}-n_{N_1}^{\rm eq}\right)+\frac{\Gamma_{\phi\to N_1N_1}}{m_\phi}\,(1+w)\,\rho_\phi+\gamma_{ii\to NN}
\nonumber\\&
\Dot{n}_{B-L}+ 3\,H\,n_{B-L}=-\langle \Gamma_N\rangle\,\left[\epsilon_{\Delta L}\,(n_{N_1}-n_{N_1}^{\rm eq})+\frac{n_{N_1}^{\rm eq}}{2\,n_\ell^{\rm eq}}\,n_{B-L}\right]
\nonumber\\&
3\,H^2\,M_P^2= \rho_\phi+\rho_R+ n_{N_1}\,E_N
\nonumber\\&
E_N^2=M_N^2 + \left(\frac{m_\phi(a)}{2}\,\frac{a_I}{a}\right)^2\,.
\end{align}
Should we take $NN \rightarrow ii$ contribution to radiation equation? It may be negligible though, but may be added as the corresponding rate is added to the BE of N.)
Here $i\in\{\Psi,\,\varphi\}$, depending on fermionic and bosonic reheating, respectively. The Yukawa coupling stems from Eq.~\eqref{eq:yukawa}, as discussed in Sec.~\ref{sec:model}. Here
\begin{align}
\langle \Gamma_N\rangle= \frac{K_1(M_N/T)}{K_2(M_N/T)}\frac{M_N}{8\pi} \left(y_N^\dagger y_N\right)_{11}\,,  
\end{align}
is the thermally averaged RHN decay width with $K_i$'s being modified Bessel functions of $i$-th kind. The equilibrium number density of any species $j$ is given by
\begin{align}
    n_j^{\rm eq}= \frac{g_j\,T^3}{2 \pi^2} \left(\frac{M_j}{T}\right)^2 K_2(M_j/T),
\end{align}
with $g_j$ representing the degrees of freedom of the $j$-species. We make the following re-definition of variables in order to solve the above set of BEQs
\begin{align}\label{eq:variable}
& \Phi=\rho_\phi\,A^{3\,(1+w)},\,R=\rho_R\,A^4,N_1=n_{N_1}\,A^3,\,\,N_{B-L}=n_{B-L}\,A^3\,.
\end{align}
With these redefined variables we express Eq.~\eqref{eq:BEQ-phiNN} as
\begin{align}
& \Phi' = -(1+w)\,\frac{\Phi\,\left(\Gamma_{\phi\to ii}+\Gamma_{\phi\to N_1N_1}\right)}{A\,H}
\nonumber\\&
R'=(1+w)\,\frac{\Phi\,\Gamma_{\phi\to ii}}{A^{3w}\,H}+\frac{\langle\Gamma_N\rangle}{H}\,\left(N_1-N_1^{\rm eq}\right)\,E_N
\nonumber\\&
N_1' = -\frac{\langle\Gamma_N\rangle}{A\,H}\,\left(N_1-N_1^{\rm eq}\right)+\frac{(1+w)}{A^{(1+3w)}}\,\frac{\Phi\,\Gamma_{\phi\to N_1N_1}}{H\,m_\phi}+\frac{A^2\,\gamma_{ii\to NN}}{H}
\nonumber\\&
N_{B-L}' = -\frac{\langle\Gamma_N\rangle}{A\,H}\,\left[\epsilon_{\Delta L}\,(N_1-N_1^{\rm eq})+\frac{N_1^{\rm eq}}{2\,N_\ell^{\rm eq}}\,N_{B-L}\right]
\nonumber\\&
3\,H^2\,M_P^2 = \frac{1}{A^3}\,\left(\frac{\Phi}{A^{3w}}+\frac{R}{A}+N_1\,E_N\right)\,,
\nonumber\\&
E_N^2=M_N^2 + \left(\frac{m_I}{2}\,A^\frac{4n-4}{n+2}\,a_I^{-3\,\frac{n-2}{n+2}}\right)^2\,,
\label{eq:BE2}
\end{align}
with $A=a/a_I$, and $n_\ell^{\rm eq}=\sum_{i=1}^3\,n_{\ell_i}^{\rm eq}$ is the sum over three lepton flavours.

The reaction rate density of the $2\to2$ processes are encoded in $\gamma_{ii\to NN}$, where \begin{align}\label{eq:gamma-22}
\gamma_{ii\to NN}=\frac{T}{32\pi^4}\,g_a g_b\,\int_{4M_N^2}^{\infty} ds\,s^{3/2}\,\sigma\left(s\right)_{ii\to NN}\,K_1\left(\frac{\sqrt{s}}{T}\right)\,,  
\end{align}
with $g_{a,b}$ being the internal degrees of freedom of the initial states. The production cross-sections read, 
\begin{align}\label{eq:cs}
& \sigma(s)_{hh\to NN}=\frac{1}{8\pi}\,\frac{\mueff^2\,\yNN^2}{(s-m_\phi^2)^2+\Gamma_\phi^2\,m_\phi^2}\,\,\left(1-\frac{4M_N^2}{s}\right)^{3/2}
\nonumber\\&
\sigma(s)_{\Psi\Psi\to NN}=\frac{s}{8\pi}\,\frac{\yeff^2\,\yNN^2}{(s-m_\phi^2)^2+\Gamma_\phi^2\,m_\phi^2}\,\left(1-\frac{4M_N^2}{s}\right)^{3/2}\,,
\end{align}
where $\Gamma_\phi$ is the total inflaton decay width, i.e., $\Gamma_\phi=\Gamma_{\phi\to ii}+\Gamma_{\phi\to NN}$. To obtain an approximate analytical expression for $\gamma$, we consider the following scenarios:
\begin{itemize}
\item [(i)] For $m_\phi\gg M_N$, considering bosonic reheating, i.e., $g_a=g_b=4$,
\begin{align}
& \gamma_{hh\to NN}\simeq 
\frac{\mueff^2\,\yNN^2}{16\pi^4}\,T
\begin{cases}
\frac{2\,T^5}{\pi\,m_\phi^4}\,, & T\ll m_\phi
\\[10pt]
\frac{m_\phi^2}{\Gamma_{\phi\to hh}}\,K_1\left(m_\phi/T\right)\,\left(1-\frac{4M_N^2}{m_\phi^2}\right)^{3/2}\,, & T\gg m_\phi
\end{cases}
\end{align}
On the other hand, for fermionic reheating ,
\begin{align}
& \gamma_{\Psi\Psi\to NN}\simeq \frac{g_a\,g_b}{256\pi^4}\,\yeff^2\,\yNN^2
\begin{cases}
\frac{512\,T^8}{\pi\,m_\phi^4}\,, & T\ll m_\phi
\\[10pt]
\frac{T\,m_\phi^4}{\Gamma_{\phi\to\Psi\Psi}}\,K_1\left(m_\phi/T\right)\,\left(1-\frac{4M_N^2}{m_\phi^2}\right)^{3/2}\,, & T\gg m_\phi
\end{cases}
\end{align}
with $g_a=g_b=72$ for quarks, $g_a=g_b=12$ for charged leptons and  $g_a=g_b=6$ for light neutrinos. 
\item [(ii)] For $m_\phi\ll M_N$, considering bosonic reheating,
\begin{align}
& \gamma_{hh\to NN}\simeq
\mueff^2\,\yNN^2
\begin{cases}
\frac{e^{-2\,M_N/T}}{32\,M_N\,\pi^4}\times\mathcal{A}\,, & T\ll M_N
\\[10pt]
\frac{3\,M_N}{64\,\pi^{9/2}}\,G_{1,3}^{3,0}\left(\frac{M_N^2}{T^2}\Big|
\begin{array}{c}
2 \\
-\frac{1}{2},-\frac{1}{2},\frac{1}{2} \\
\end{array}
\right)\,, & T\gg M_N\,,
\end{cases}
\end{align}
where $\mathcal{A}=4\,M_N^3\, e^{\frac{2 M_N}{T}}\, \text{E}_i\left(-\frac{2 M_N}{T}\right)+2 M_N^2\,T \left[1-T/(2M_N)+\left(T/\sqrt{2}\,M_N\right)^2\right]$, $\text{E}_i[z]$ being the exponential integral function and $G$ represents the Meijer G-function. On the other hand, for fermionic reheating ,
\begin{align}
& \gamma_{\Psi\Psi\to NN}\simeq \frac{g_a\,g_b}{256\,\pi^5}\,\yeff^2\,\yNN^2
\begin{cases}
T\cdot\mathcal{B}\,, & T\ll M_N
\\[10pt]
4\,T^4\,, & T\gg M_N\,,
\end{cases}
\end{align}
where $\mathcal{B}=\frac{8 \pi\,M_N^4}{T}\,\text{E}_i\left(-\frac{2 M_N}{T}\right)+4 \pi\,M_N^3\,e^{-\frac{2 M_N}{T}} \left[1-T/(2\,M_N) +\left(T/\sqrt{2}\,M_N\right)^2\right]$.
\end{itemize}
Here, we have ignored the
masses of the initial states, which are the SM fields. This is a reasonable approximation, where reheating happens at a very high temperature, much before the electroweak symmetry is broken. Although, depending on the RHN mass, $m_\phi> M_N$ can be true at the very beginning of reheating, however, at some point $m_\phi<M_N$ is possible for $n>2$. Typically, for $a>a_\star$ [cf.Eq.~\eqref{eq:ast}], the inflaton decay is no more viable, but still production via inflaton-mediation is possible.

Sphaleron interactions are in equilibrium in the temperature range between $\sim$ 100 GeV and $10^{12}$ GeV, and they convert a fraction of a non-zero $B-L$ asymmetry into a baryon asymmetry via
\begin{align}
Y_B\simeq a_{\rm sph}\,Y_{B-L}=\frac{8\,N_F+4\,N_H}{22\,N_F+13\,N_H}\,Y_{B-L}\,,    
\end{align}
where $N_F$ is the number of fermion generations and $N_H$ is the number of Higgs doublets, which in our case: $N_F = 3,\,N_H = 1$ and $a_{\rm sph}\simeq 28/79$. In leptogenesis, where purely a lepton asymmetry is generated, $B-L=-L$. This is converted into the baryon asymmetry via sphaleron transition~\cite{Buchmuller:2004nz}. Finally, the observed baryon asymmetry of the Universe is given by $Y_B^0 \simeq 8.75\times 10^{-11}$.
\begin{figure}[htb!]
    \centering
    \includegraphics[scale=0.375]{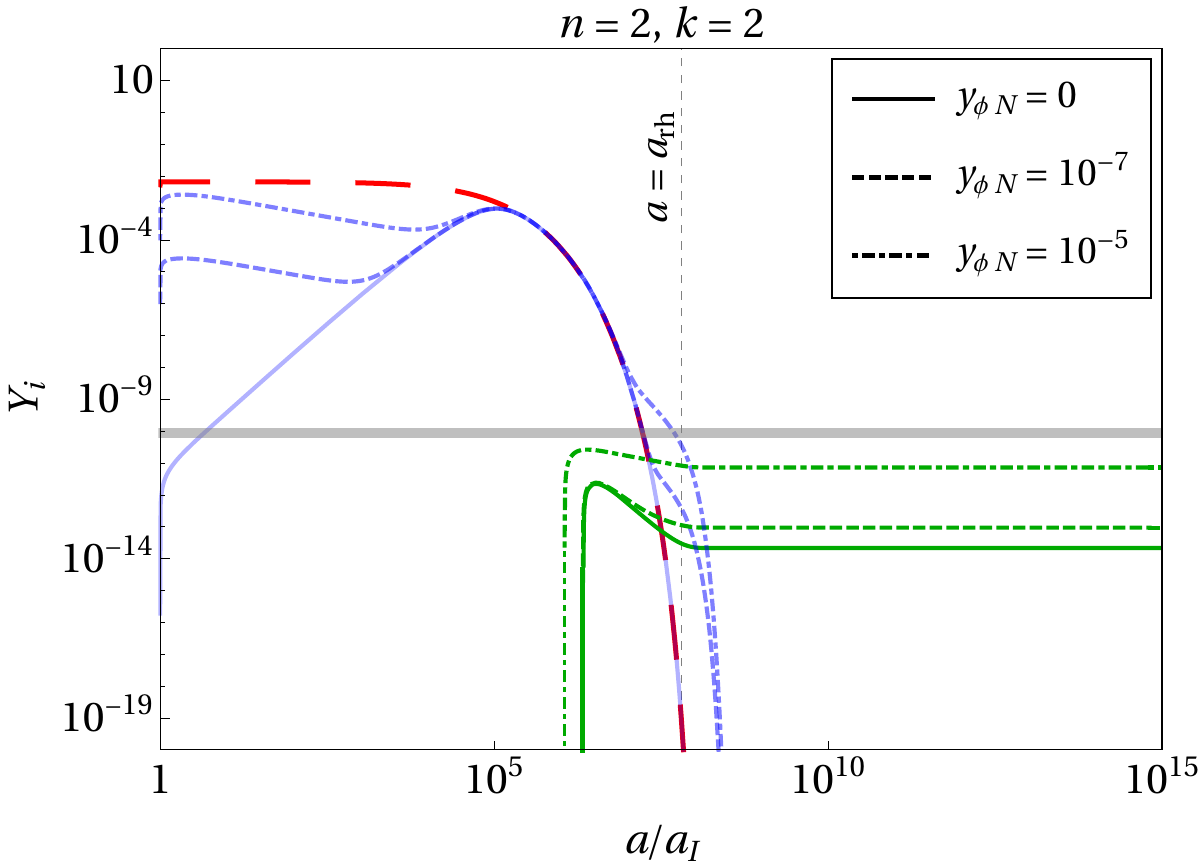}~\includegraphics[scale=0.375]{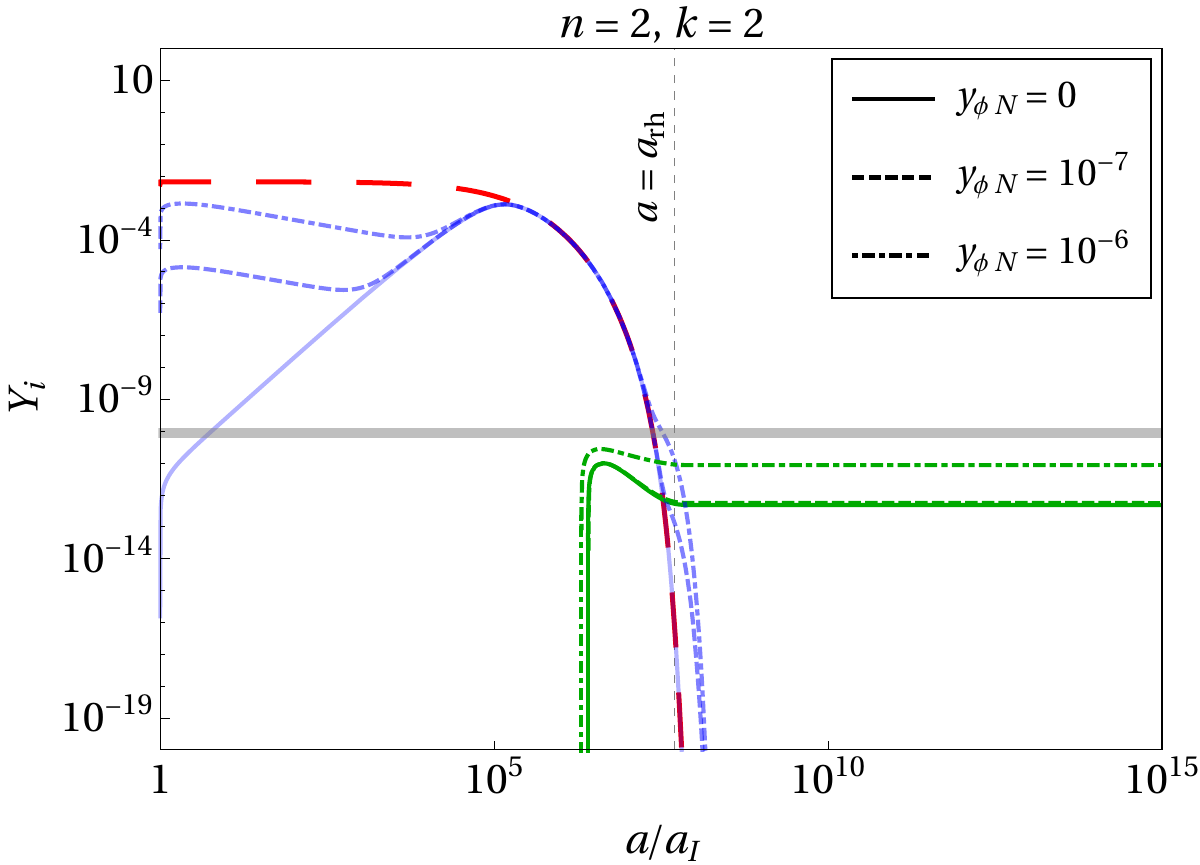}\\[10pt]
    \includegraphics[scale=0.375]{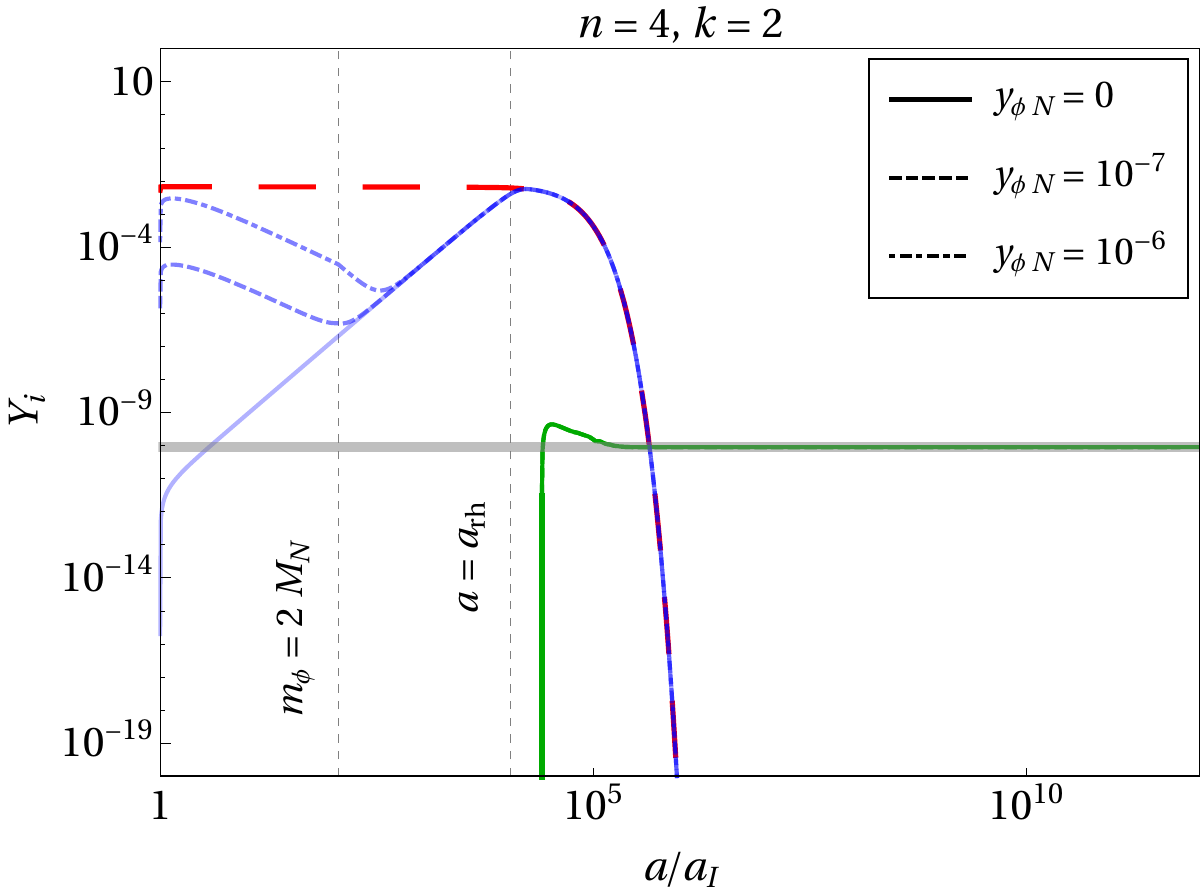}~\includegraphics[scale=0.375]{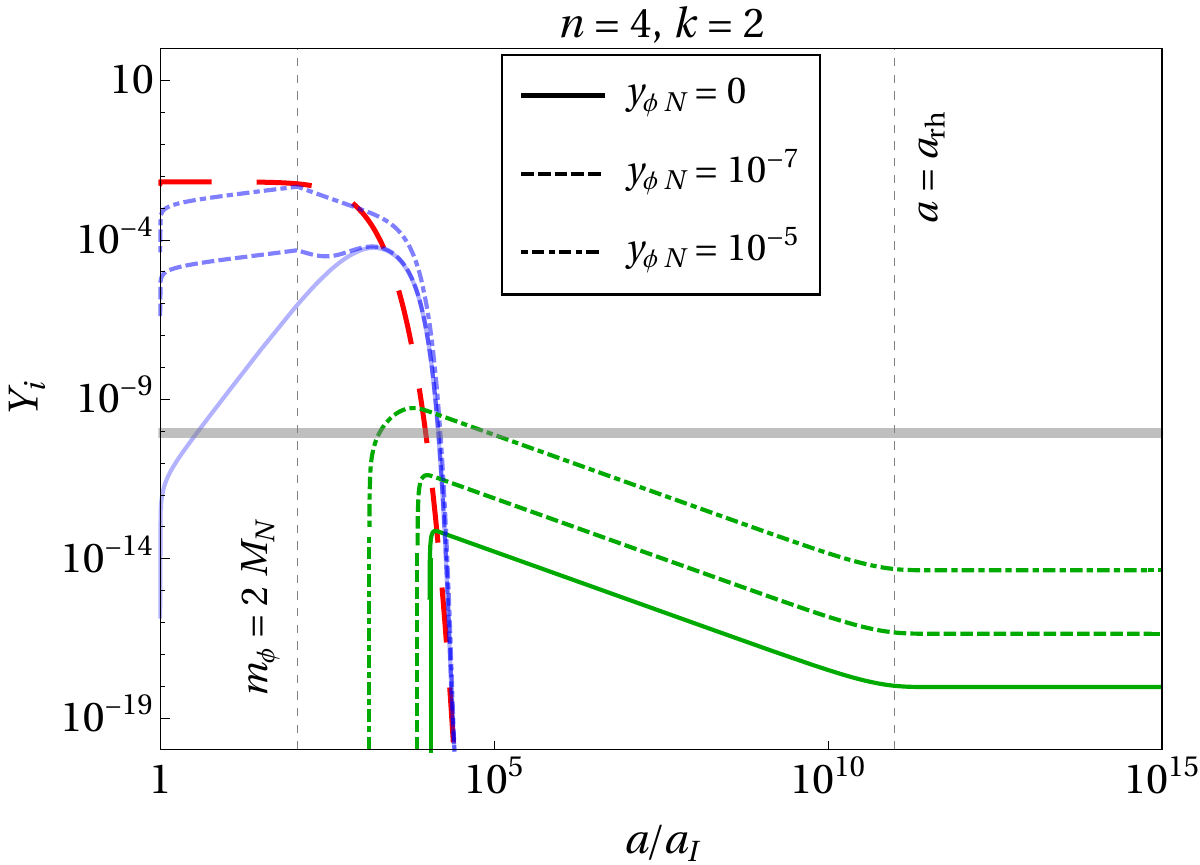}\\[10pt]
    \includegraphics[scale=0.375]{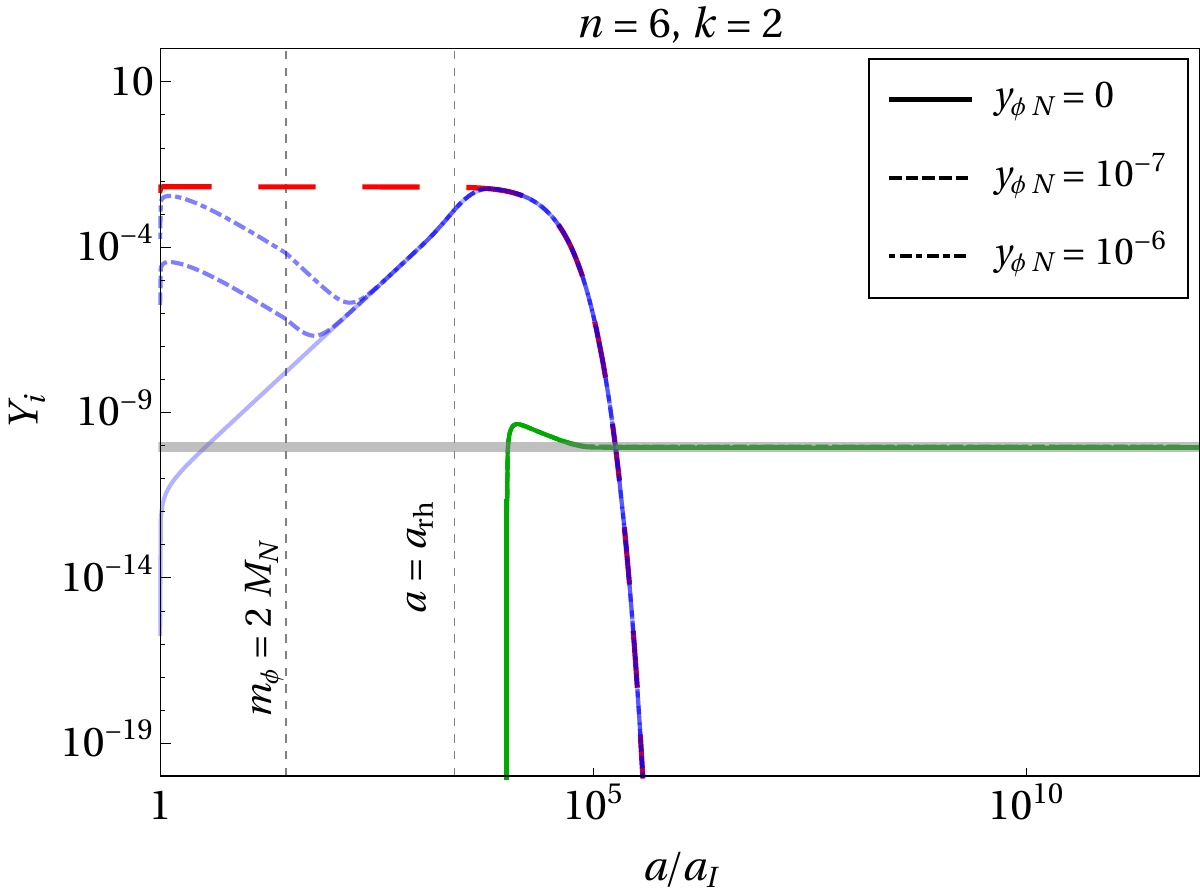}~\includegraphics[scale=0.375]{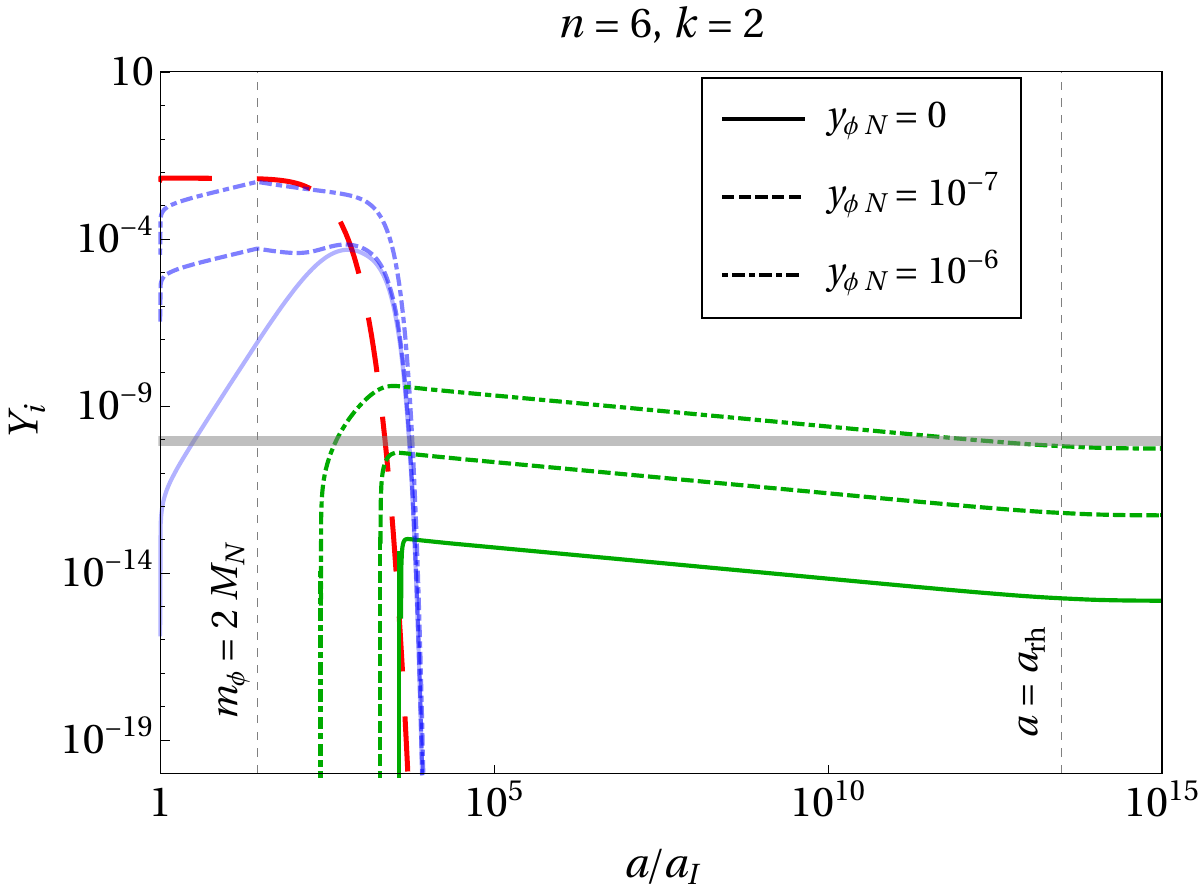}
    \caption{Left column: bosonic reheating scenario. We show the evolution of yield of RHN number density (blue) and $B-L$ number density (green), with the scale factor for $n=2$ (top), $n=4$ (middle) and $n=6$ (bottom). Here we choose $\meff= 10^{10}$ GeV. Right column: fermionic reheating scenario. Here we choose $\yeff=10^{-5}$. In all cases we fix $k=2$ and $M_N=10^{11}$ GeV. The reheating temperatures are same as those in Figs.~\ref{fig:rho-T1-b} and ~\ref{fig:rho-T1-f}. The thick dashed red curve corresponds to equilibrium number density. Different curves correspond to different choices of $\yNN$ as mentioned in the plot legend.}
    \label{fig:yield}
\end{figure}
\begin{figure}[htb!]
    \centering    \includegraphics[scale=0.375]{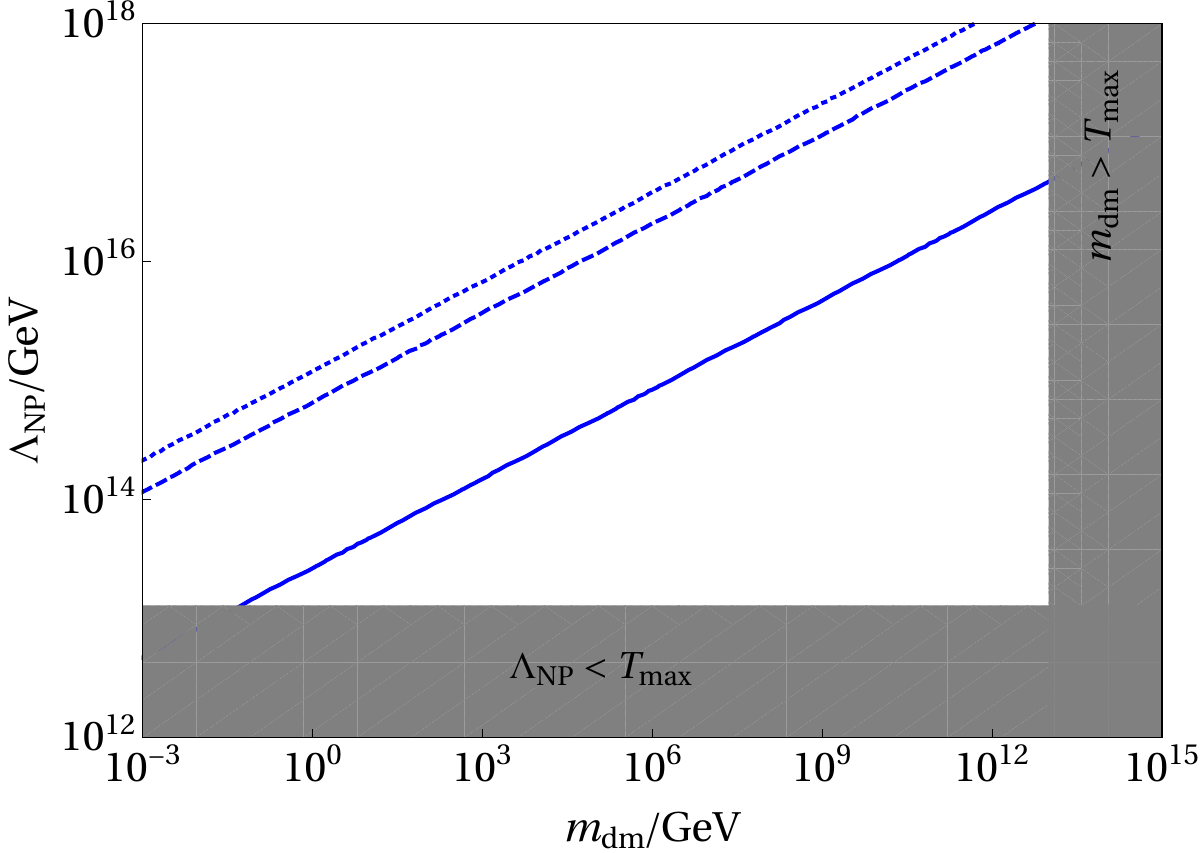}~\includegraphics[scale=0.375]{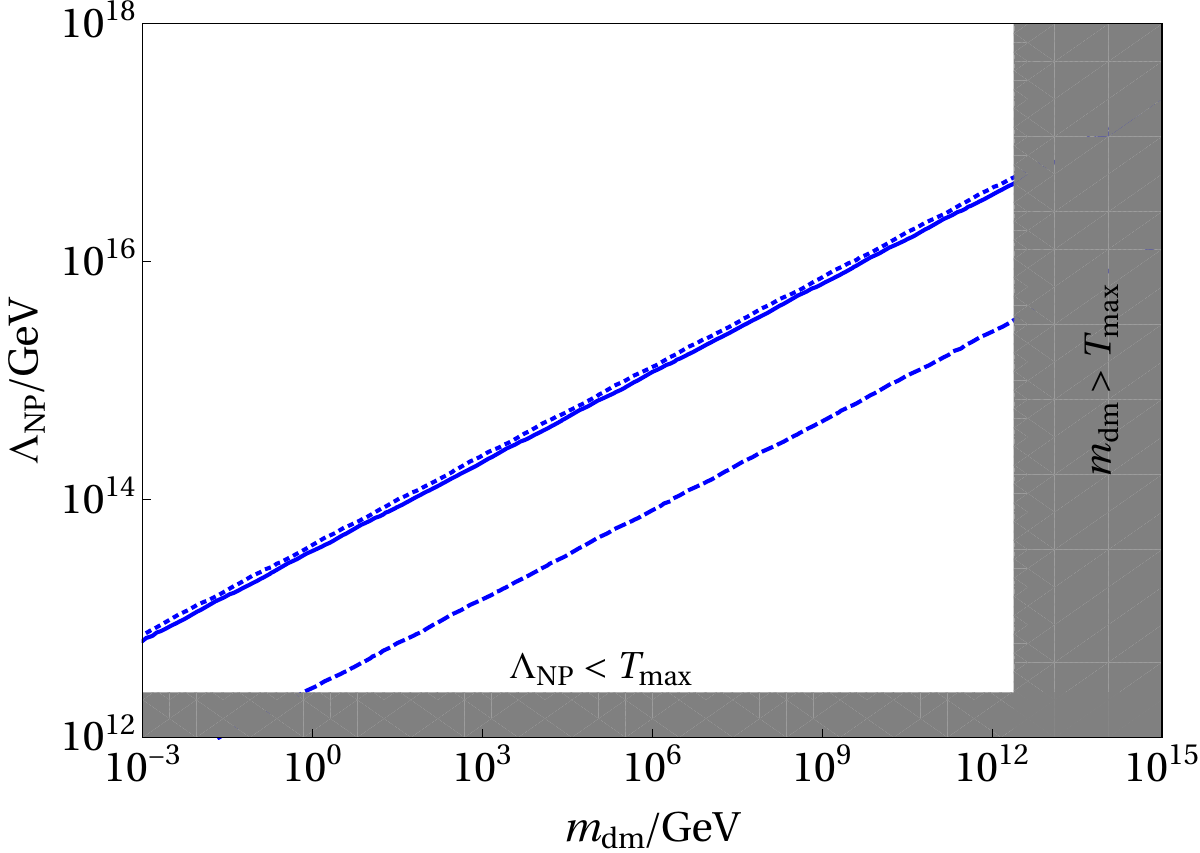}
    \caption{Contours satisfying observed DM abundance for $n=2,\,4,\,6$, shown via solid, dashed and dotted curves respectively for bosonic (left) and fermionic (right) reheating scenarios. The inflaton couplings correspond to those which give rise to right baryon asymmetry. The shaded regions are forbidden from max DM mass produced during reheating: $\mdm\lesssim\Tmax$ and validity of the effective DM-SM description: $\lNP>\Tmax$. We choose $\mueff=10^{8}$ GeV and $\yeff=4\times10^{-5}$, and the corresponding $\Trh$ can be read off from Fig.~\ref{fig:rho-T1-b} and \ref{fig:rho-T1-f}. In all cases we choose $k=2$.}
    \label{fig:dm}
\end{figure}

\subsection*{Dark matter production}
Considering DM to be produced entirely from the SM bath, the DM number can then be tracked by solving the Boltzmann equation for its number density $\ndm$
\begin{equation} \label{eq:BEDM0}
    \frac{d\ndm}{dt} + 3\, H\, \ndm = \gamma\,,
\end{equation}
where $\gamma$ is the DM production reaction rate density, which we parametrize as~\cite{Elahi:2014fsa,Bernal:2019mhf,Kaneta:2019zgw,Barman:2023ktz}\footnote{Here we consider DM production only from the SM bath, while interactions involving a pair of RHNs and a pair of DM or a pair of SM fields are absent by construction.}
\begin{align}
& \gamma=\frac{T^{k+6}}{\Lambda_{\rm NP}^{k+2}}\,.  
\end{align}
Here $k=2\,(d-5)$, where $d$ is the dimension of the relevant effective DM-SM operator ($d \geq 5$). The mass scale $\lNP$ is identified with
the beyond the SM scale of the microscopic model under consideration. Note that, this effective description is valid for the duration of reheating provided
that $\lNP\gtrsim\Tmax$. As the SM entropy is not conserved during reheating due to the decay of the inflaton in SM particles, it is convenient to introduce a comoving number density $\Ndm \equiv \ndm\, a^3$, and therefore Eq.~\eqref{eq:BEDM0} can be rewritten as
\begin{equation} \label{eq:BEDM}
    \frac{d\Ndm}{da} = \frac{a^2\,\gamma}{H}\,,
\end{equation}
which has to be numerically solved together with Eqs.~\eqref{eq:drhodt} and~\eqref{eq:BEQ-phiNN}, taking the initial condition $\Ndm(a_I) = 0$. Now, since the Hubble parameter is dominated by inflaton energy density during reheating, hence using Eq.~\eqref{eq:Hubble} we can obtain approximate analytical expression for DM yield. We note there are two possible scenarios that may appear here:
\begin{itemize}
\item For $\mdm\ll\Trh$, the DM is produced at the {\it end of} reheating, with corresponding yield
\begin{align}\label{eq:ydm1}
& Y(\arh)\simeq\frac{45\,\sqrt{10}\,(n+2)}{\pi^3\,\gss(\Trh)^{3/2}}\,\frac{M_P\,\Trh^{k+1}}{\lNP^{k+2}}
\nonumber\\&\times
\begin{cases}
\frac{1}{k-n\,(k+4)+10}\,\left[1-\left(a_I/\arh\right)^{\frac{3\,(k+10-n\,(k+4)}{2n+4}}\right]\,, & \text{fermionic reheating}
\\[10pt]
\frac{1}{2\,n-k-2}\,\left[1-\left(a_I/\arh\right)^\frac{6n-3k-6}{2n+4}\right]\,, & \text{bosonic reheating}\,.
\end{cases}
\end{align}
where $a_I/\arh=\left[H(\arh)/H_I\right]^\frac{n+2}{3\,n}$, following Eq.~\eqref{eq:rpsol}, $H_I$ being the Hubble parameter at the end of inflation.
\item For $\Trh<\mdm\ll\Tmax$, DM is produced {\it during} reheating. In this case, the DM number density can be estimated by integrating Eq.~\eqref{eq:BEDM} from the $a_I$ to $\adm=a(T=\mdm)$, where $\adm = \arh\times\left(\Trh/\mdm\right)^{1/\xi}$, as can be obtained using Eq.~\eqref{eq:Tevol}. In this case, DM yield at $a=\adm$ reads
\begin{align}\label{eq:ydm2}
& Y(\adm)\simeq\frac{45\,\sqrt{10}\,(n+2)}{\pi^3\,\gss^{3/2}(\mdm)}\,\frac{M_P\,\Trh^{k+4}}{\mdm^3\,\lNP^{k+2}}
\nonumber\\&\times
\begin{cases}
\left(\Trh/\mdm\right)^\frac{6+k-n\,(k+4)}{n-1}\,\frac{1}{k-n\,(k+4)+10}\,\left[1-\left(a_I/\adm\right)^{\frac{3\,(k+10-n\,(k+4)}{2n+4}}\right]\,, & \text{fermionic reheating}
\\[10pt]
\left(\Trh/\mdm\right)^{2n-k-6}\,\frac{1}{2\,n-k-2}\,\left[1-\left(a_I/\adm\right)^\frac{6n-3k-6}{2n+4}\right]\,, & \text{bosonic reheating}\,.
\end{cases}
\end{align}
The final yield at the end of reheating, in this case,
\begin{align}
& Y(\arh) = Y(\adm)\, \left(\frac{\adm}{\arh} \right)^3\frac{s(\adm)}{s(\arh)} \simeq Y(\adm)\,\left(\frac{\mdm}{\Trh}\right)^3\,\left(\frac{\adm}{\arh}\right)^3\,,
\end{align}
up to some order of the ratio of $\gss$. Here, $a_I/\adm=\left[H(\arh)/H_I\right]^\frac{n+2}{3n}\times\left(\arh/\adm\right)$ with $\arh/\adm=\left(\mdm/\Trh\right)^{1/\xi}$, following Eq.~\eqref{eq:rpsol} and \eqref{eq:Tevol}.
\end{itemize}
To fit the whole observed DM relic density, it is required that
\begin{equation} \label{eq:obsyield}
    Y_0\, \mdm = \Omega h^2 \, \frac{1}{s_0}\,\frac{\rho_c}{h^2} \simeq \mathcal{C}_{\rm rel}\,,
\end{equation}
where $Y_0 \equiv Y(\arh)$ (assuming no entropy injection from the end of reheating till today) and $Y(T) \equiv \ndm(T)/s(T)$, with $s(T) = \frac{2\pi^2}{45}\, \gss(T)\, T^3$ being the SM entropy density and $\gss(T)$ being the number of relativistic degrees of freedom contributing to the SM entropy. Furthermore, $\rho_c \simeq 1.05 \times 10^{-5}\, h^2$~GeV/cm$^3$ is the critical energy density, $s_0\simeq 2.69 \times 10^3$~cm$^{-3}$ the present entropy density~\cite{ParticleDataGroup:2022pth}, and $\Omega h^2 \simeq 0.12$ the observed abundance of DM relics~\cite{Planck:2018vyg}, with $\mathcal{C}_{\rm rel}=4.3 \times 10^{-10}$ GeV. The DM yield is crucially dependent on $\Trh$ and $\Tmax$, which are determined by the inflaton coupling to bosonic or fermionic final states. Since these couplings also decide the generation of baryon asymmetry, therefore we have a scenario where both DM and baryon abundance are controlled by a set of common parameters, namely, the inflaton coupling to SM fields $(\mueff,\,\yeff)$ and the shape of the inflaton potential during reheating (decided by $n$), for a given DM mass and the scale of DM-SM effective interaction.

In Fig.~\ref{fig:yield} we show the evolution of yield of RHN number and $B-L$, with the scale factor. As expected, both the yields yield grows with the scale factor, eventually becoming saturated in the asymptotic limit. Now, the RHNs are sourced from the thermal bath, as well as from the inflaton decay and the inflaton-mediated scattering. We have found, the scattering channels are sub-dominant for our choice of the parameters. For $n=2$, as we see, a larger $\yNN$ results in larger final asymmetry since the $\phi\to NN$ decay always contributes to RHN production. Notice that, in this case the RHN yield does not follow equilibrium distribution for $\yNN>0$. As expected, the $B-L$ yield becomes constant the moment the rate of RHN decay surpasses its production. For $n=4$, in case of bosonic reheating, the final asymmetry is indistinguishable for different $\yNN$ values. The same is also true for $n=6$. This is because in these two cases the asymmetry is produced {\it after} reheating, during radiation domination [cf. Fig.~\ref{fig:rho-T1-b}]. Consequently, the final asymmetry is entirely due to the thermal bath, and is insensitive to the choice of $\yNN$. This situation can be changed by choosing a smaller $\mueff$ for $n=4,\,6$, that will result in longer lifetime of the inflaton resulting in longer period of reheating. This feature, however, does not happen for fermionic reheating scenario. In this case, for $n=4$ and $n=6$, a larger $\yNN$ still results in larger asymptotic yield. This is due to the fact that for fermionic reheating scenario, the asymmetry is produced {\it during} reheating [cf. Fig.~\ref{fig:rho-T1-f}]. But since reheating takes place for a longer period of time for $n>2$, this results in entropy dilution to the final asymmetry. It is worth noting here, for $n=4$, $n_{B-L}/s\sim n_N/s\sim a^{-3/4}$, while for $n=6$, $n_{B-L}/s\sim a^{-3/16}$. As a result, we see, for $n=4$, the final asymmetry is under produced compared to $n=6$ scenario. However, it is still possible to achieve right asymmetry for $n=4$ by tuning the Casas-Ibarra parameters.

The dependence of baryon asymmetry on the inflaton-RHN coupling $\yNN$ is illustrated in Fig.~\ref{fig:asym}, for both bosonic (left panel) and fermionic (right panel) reheating scenario. Here we have fixed $\text{Re}[z]=2$, $\text{Im}[z]=0.3$ for fermionic case, while for bosonic, $\text{Re}[z]=5.5$, $\text{Im}[z]=0.55$. For $n=2$, since the two reheating scenarios are identical, hence we see $Y_B$ has exactly similar behavior. The effect of $\yNN$ becomes pronounced for $\yNN\gtrsim 10^{-5}$. The $\phi\to NN$ production channel results in a significant rise in $N_1$ abundance, that in turn leads to a larger production of the asymmetry. In case of bosonic reheating, since for both $n=4$ and $n=6$, the asymmetry is produced post-reheating, hence the asymmetry is approximately independent of $\yNN$. The effect of this additional channel (on top of thermal bath) again shows up for the fermionic reheating scenario for both $n=4,\,6$. However, since now the asymmetry is produced entirely during reheating, also reheating lasts longer compared to $n=2$ case [cf. Fig.~\ref{fig:yield}], hence increase in $\yNN$ results in a significant production of asymmetry. Note that, for $n=4$, the resulting asymmetry is much smaller compared to $n=2,\,6$, a feature that we have already noticed in Fig.~\ref{fig:yield}.

The viable parameter space satisfying observed DM abundance is shown in Fig.~\ref{fig:dm}, where in the left panel we consider bosonic reheating, while the right panel considers fermionic reheating scenario. Here we have fixed $k=2$, such that the non-renormalizable DM-SM interaction is of dimension five $(d=5)$. For a fixed $n$, we vary the effective scale of interaction $\lNP$ and the DM mass in order to obtain the right abundance. The nature of the contours satisfying correct relic density can be understood, for example, from Eq.~\eqref{eq:ydm1} by noting that for $\mdm\ll\Trh$,
\begin{align}
\frac{\mdm}{\lNP^{k+2}}\propto \frac{1}{M_P}
\begin{cases}
\frac{k-n\,(k+4)+10}{\Trh^{k+1}}\,, & \text{fermionic reheating}   \\[10pt]
\frac{2n-k-2}{\Trh^{k+1}}\,, & \text{bosonic reheating}\,,
\end{cases}
\end{align}
which shows for a fixed $n,\,k$ and inflaton coupling (which in turn fixes $\Trh$), heavier DM requires larger $\lNP$-suppression to provide correct abundance. The same can also be shown for $\Trh<\mdm\ll\Tmax$, from Eq.~\eqref{eq:ydm2}. Here we choose inflaton couplings correspond to right baryon asymmetry, following Fig.~\ref{fig:asym}. It is therefore the same set of couplings that are responsible for the simultaneous generation of right baryon asymmetry and DM abundance. 

\section{Spectrum of inflationary gravitational wave}
\label{sec:pgw}
The spectrum of GWs is described in terms of the fraction of their energy density per logarithmic wavenumber (frequency) $k$ interval
\begin{equation} \label{eq:omega-gw}
    \ogw(a,k) = \frac{1}{\rho_c}\,\frac{d\rGW}{d\log k}\,,
\end{equation}
normalized to the critical density $\rho_c=3\,H^2\, M_P^2$. This expression can be recast as~\cite{Maggiore:1999vm, Watanabe:2006qe, Saikawa:2018rcs, Caprini:2018mtu}
\begin{equation} \label{eq:ogw-k}
    \ogw(a\,,k) = \frac{1}{12} \left(\frac{k}{a\, H(a)}\right)^2 \mathcal{P}_{T,\text{prim}}\, \mathcal{T}(a,k)\,,
\end{equation}
where the primordial tensor power spectrum $\mathcal{P}_{T,\text{prim}}$ can be parametrized as~\cite{Saikawa:2018rcs, Caprini:2018mtu}
\begin{equation}
    \mathcal{P}_{T,\text{prim}} = r\, \mathcal{P}_{\zeta}(k_\star) \left(\frac{k}{k_\star}\right)^{n_T}\,,
\end{equation}
with $k_\star = 0.05~\text{Mpc}^{-1}$ being the Planck pivot scale,  $\mathcal{P}_{\zeta}(k_\star)\simeq 2.1\times 10^{-9}$~\cite{Planck:2018jri} corresponding power spectrum of the scalar perturbation at the pivot scale, $r$ is the tensor-to-scalar ratio and $n_T$ represents the tensor spectral index. In single-field inflationary models, $n_T \simeq -r/8$. Considering the current bound on $r<0.036$~\cite{BICEP:2021xfz}, we set $n_T =0$ throughout our analysis, which corresponds to a scale-invariant primordial spectrum. In Eq.~\eqref{eq:ogw-k} the transfer function $\mathcal{T}(a,\,k)$ connects primordial mode functions with mode functions at some later time as~\cite{Boyle:2005se, Saikawa:2018rcs}
\begin{equation}\label{eq:T-fun}
    \mathcal{T}(a,\,k) = \frac12 \left(\frac{\ahc}{a}\right)^2,
\end{equation}
where the prefactor 1/2 appears due to oscillation-averaging the tensor mode functions~\cite{Saikawa:2018rcs, Figueroa:2018twl, Choi:2021lcn} and  $\ahc$ corresponds to the scale factor at the horizon crossing (labeled as ``hc") defined as $\ahc\,  H(\ahc) = k$. Different $k$ modes might reenter the horizon at different epochs, {\it viz,} during radiation domination or during inflaton domination (reheating). From Eq.~\eqref{eq:ogw-k}, one can obtain the spectral GW energy density at present as,
\begin{equation} \label{eq:ogw-0}
    \ogw(k) \equiv \ogw(a_0, k) = \frac{1}{24}\left(\frac{k}{a_0\, H_0}\right)^2 \mathcal{P}_{T,\text{prim}} \left(\frac{\ahc}{a_0}\right)^2,
\end{equation}
where $H_0$ is the Hubble rate at the present epoch. Taking into account the reentry of modes at different epochs, the full GW spectral energy density can be expressed in piece-wise form as
\begin{align} \label{eq:ogw}
    \ogw(\ahc) &\simeq \Omega_{\gamma}^{(0)}\, \frac{\mathcal{P}_{T,\text{prim}}}{24}\, \frac{\gs(T_\text{hc})}{2} \left(\frac{\gss(T_0)}{\gss(T_\text{hc})}\right)^\frac43 \nonumber\\
    & \qquad\qquad \times
    \begin{cases}
        \frac{\gs(\Trh)}{\gs(T_\text{hc})} \left(\frac{\gss(T_\text{hc})}{\gss(\Trh)}\right)^\frac43 \left(\frac{\arh}{\ahc}\right)^{\frac{2(n-4)}{n+2}} &\text{for } a_I < \ahc \leq \arh\,,\\
        1 &\text{for }\arh \leq \ahc \leq a_{\rm eq}\,,
    \end{cases}
\end{align}
where $\Omega_{\gamma}^{(0)} \equiv \rho_{\gamma,0}/\rho_c=2.47\times 10^{-5}\,h^{-2}$~\cite{Planck:2018vyg} is the fraction of the energy density of photons at the present epoch and $\Thc$ corresponds to the temperature at which the corresponding mode reenters the horizon. Note that, for $\arh\leq\ahc\leq \aeq$, we have assumed conservation of entropy from the moment of horizon crossing until today, implying $T\propto a^{-1}\,\gss^{-1/3}$, with $a = \aeq$ being the scale factor at the matter-radiation equality (MRE) at $T\equiv\Teq \simeq 0.7$~eV. Corresponding GW frequency $f$ can be expressed as
\begin{align}\label{eq:fre}
    f(\ahc) \equiv \frac{k}{2\pi\, a_0} = \mathcal{F}(\gs,\,\gss)\,\frac{T_0\,\sqrt{H_I\,M_P}}{M_P}\,\sqrt{h_r}
    \times
    \begin{cases}
        \left(\frac{\arh}{\ahc}\right)^\frac{2(n-1)}{n+2} &\text{for } a_I < \ahc \leq \arh\,,\\
        \frac{\arh}{\ahc} &\text{for }\arh \leq \ahc \leq a_{\rm eq}\,,
    \end{cases}
\end{align}
where $T_0 \simeq 2.3 \times 10^{-13}$~GeV~\cite{ParticleDataGroup:2022pth}, $\mathcal{F}(\gs,\,\gss)=\frac16 \sqrt{\frac{\gs(\Trh)}{10}} \left(\frac{\gss(T_0)}{\gss(\Trh)}\right)^\frac13\,\left(\frac{90}{\pi^2\,\gs(\Trh)}\right)^{1/4}$and $h_r\equiv H(\arh)/H_I$ is the ratio of Hubble parameter at the end of reheating to that at the end of inflation. This parameter is a function of the reheating temperature $\Trh$, which in turn depends on the inflaton coupling. Different choices of $h_r$, for a fixed $H_I$, therefore corresponds to different $\mueff$ or $\yeff$, which then can be associated with the scale of leptogenesis and the DM mass, following our discussion in Sec.~\ref{sec:cogenesis}. 

We can recast the GW spectral energy density as a function of its frequency, using Eqs.~\eqref{eq:ogw} and~\eqref{eq:fre} as
\begin{align}\label{eq:ogw2} 
    \ogw(f) &\simeq \Omega_{\gamma}^{(0)}\, \frac{\mathcal{P}_{T,\text{prim}}}{24}\, \frac{\gs(T_\text{hc})}{2} \left(\frac{\gss(T_0)}{\gss(T_\text{hc})}\right)^\frac43 \nonumber
    \\
    &\times
    \begin{cases}
        \frac{\gs(\Trh)}{\gs(T_\text{hc})} \left(\frac{\gss(T_\text{hc})}{\gss(\Trh)}\right)^\frac43 
 \left(\frac{f}{\frh}\right)^\frac{n-4}{n-1} &\text{for } \frh \leq f < f_\text{max}\,,\\
        1 &\text{for } f_\text{eq} \leq f \leq \frh\,,
    \end{cases}
\end{align}
with $f_\text{eq} \equiv f(\aeq)$ and
\begin{equation}
f_{\rm rh}\equiv f(\arh)=\mathcal{F}(\gs,\,\gss)\,\frac{T_0}{M_P}\,\sqrt{M_P\,H_I}\,\sqrt{h_r}\,.
\end{equation}
\begin{figure}[htb!]
    \centering
    \includegraphics[scale=0.6]{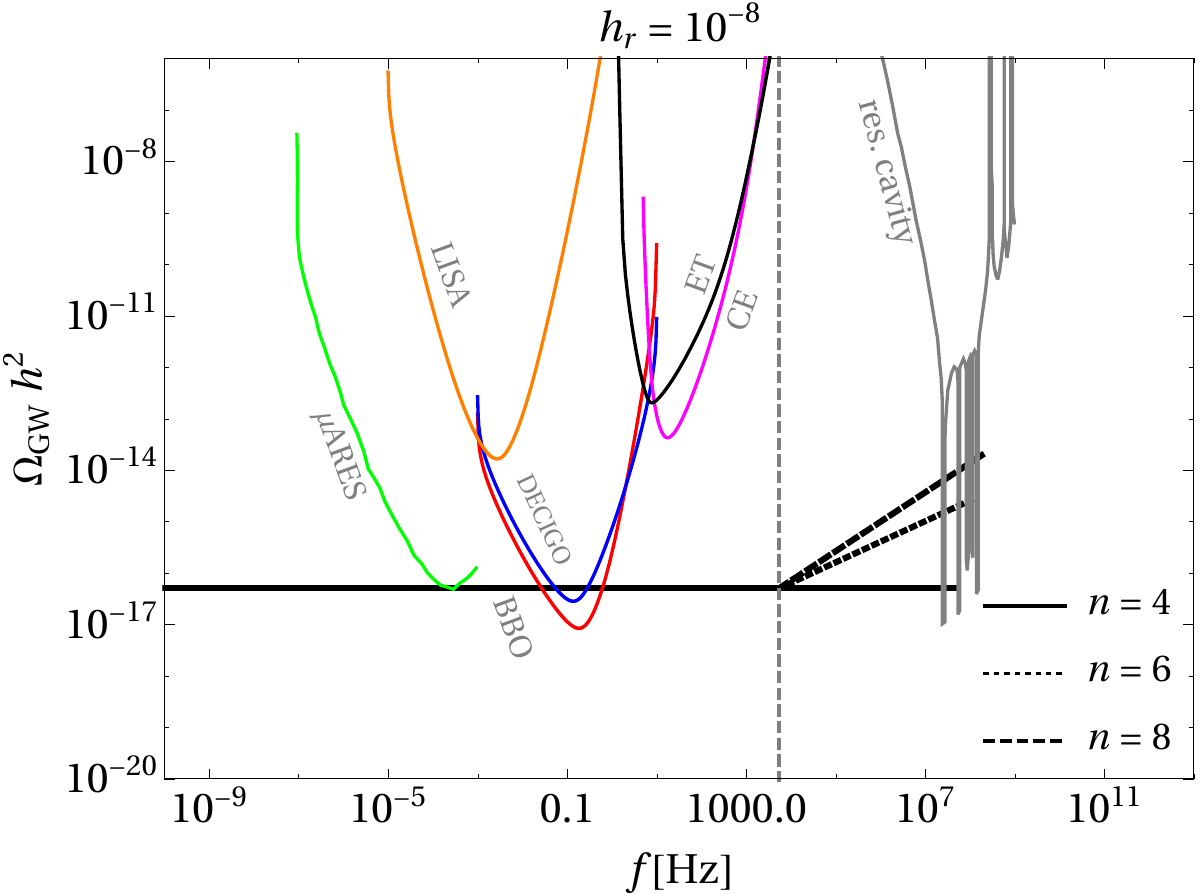}
    \\[10pt]    \includegraphics[scale=0.6]{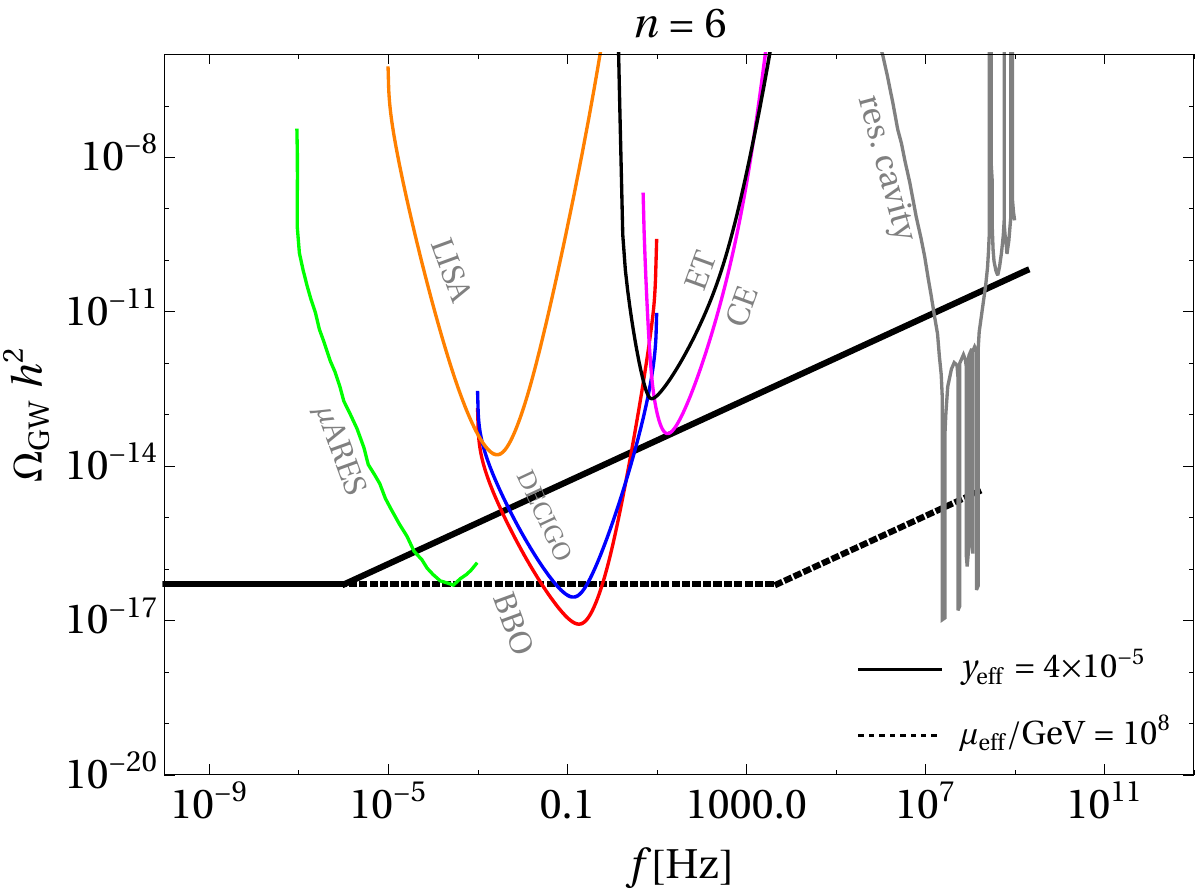}
    \caption{Top: Inflationary primordial gravitational wave spectrum as a function of frequency, for different choices of $n$, shown via different patterns for $h_r\equiv H(\arh)/H_I=10^{-8}$. Bottom: Same as top, but for a fixed $n=6$, with different choices of couplings that correspond to right baryon asymmetry and DM abundance, following Fig.~\ref{fig:asym} and \ref{fig:dm}. The scale of inflation $H_I=4.4\times 10^{13}$ GeV is fixed. We project future sensitivity from several GW detectors.}
    \label{fig:pgw}
\end{figure}
There exist an upper cut-off on GW frequency corresponding to the modes that re-entered the horizon right after the end of inflation, i.e., $\ahc =a_I$,
\begin{equation}
    f_\text{max} \equiv f(a_I) = \frac{H_I}{2\,\pi}\, \frac{a_I}{a_0}=\frac{T_0}{2\,\pi}\,\left(\frac{\pi^2\,\gs(\Trh)}{90}\right)^{1/4}\,\left(\frac{\gss(T_0)}{\gss(\Trh)}\right)^\frac13 \frac{a_I}{\arh}\,\sqrt{\frac{H_I}{M_P}}\, h_r^{-1/2} \,.
\end{equation}
From Eq.~\eqref{eq:ogw2} one can see that modes that reenter the horizon after reheating (that is $f  < \frh$) keep the same dependence with the frequency as the original primordial spectrum $\mathcal{P}_{T,\text{prim}}$. However, if they reenter during reheating ($\frh < f < \fmax$), $\ogw(f)$ gets a boost $\propto f^{\frac{n-4}{n-1}}$, which implies that the spectrum becomes blue tilted with respect to the primordial one. 

In Fig.~\ref{fig:pgw} we show GW spectral energy density at present epoch, as a function of the frequency $f$. In the top panel, for illustration, we fix $h_r=10^{-6}$, while choose different values of $n$. This implies, for different $n$, we have chosen the inflaton-SM couplings in a way that they give rise to the same $H(\arh)$, or equivalently, $\Trh$ for both bosonic and fermionic reheating scenarios. We also fix $H_I\simeq 4.4\times 10^{13}$ GeV, corresponding to the maximum value for the tensor-to-scalar ratio $r = 0.035$~\cite{BICEP:2021xfz}.As explained before, for $n=4$ we see a scale-invariant spectrum following Eq.~\eqref{eq:ogw}, while the spectrum is blue-tilted for $n>4$. For a given $h_r$ and $H_I$, as $H(\arh)$ is also fixed, hence we see for all curves $f(\arh)\simeq 5.4\times 10^4$ Hz is also constant.

In the bottom panel we fix $n=6$ as that results in a blue-tilted detectable GW spectrum. We now choose two benchmark values of the couplings $\{\yeff,\,\mueff\}=\{4\times 10^{-5},\,10^8\,\text{GeV}\}$, corresponding to fermionic and bosonic reheating, respectively, with corresponding reheating temperatures of $\Trh\simeq\{5\times 10^{11},\,10^7\}$ GeV. These choices of couplings can give rise to observed baryon asymmetry for $n=6$ for $M_N=10^{11}$ GeV, as evident from Fig.~\ref{fig:asym}. We once again point out that, while for fermionic reheating we require $\yNN>0$ to produce the observed asymmetry, for bosonic case, it is possible to have $\yNN=0$.

For all such couplings it is also possible to choose $\mdm$ and $\lNP$ that can provide right DM abundance, as seen from Fig.~\ref{fig:dm}. We see,  The reason being, a smaller $\yeff$ results in a lower $\Trh$, hence a lower $H(\arh)$ and therefore a larger $h_r$. As a result, corresponding $\frh$ is lower, or in other words, the stiff period lasts longer resulting in too much blue-tilt for the spectrum. For bosonic reheating scenario, on the other hand, reheating gets completed at an  early stage [cf. Fig.~\ref{fig:rho-T1-b}], resulting in a very high $\Trh$ (and hence $H(\arh)$). Consequently, corresponding $\frh$ is higher, i.e., the stiff epoch ends earlier, and hence the corresponding spectrum has a blue-tilt at a high frequency. We project sensitivity curves from several future experiments, for example, the Big Bang Observer (BBO)~\cite{Crowder:2005nr, Corbin:2005ny}, ultimate DECIGO (uDECIGO)~\cite{Seto:2001qf, Kudoh:2005as}, LISA~\cite{LISA:2017pwj}, $\mu$Ares~\cite{Sesana:2019vho}, the cosmic explorer (CE)~\cite{Reitze:2019iox}, the Einstein Telescope (ET)~\cite{Hild:2010id, Punturo:2010zz, Sathyaprakash:2012jk, Maggiore:2019uih}, and from resonant cavities~\cite{Herman:2022fau}. 
\section{Conclusion}
\label{sec:concl}
In this work we have discussed the production of baryon asymmetry, together with dark matter (DM) during the period of reheating after inflation, when the energy density of the Universe is dominated by a single scalar field, the inflaton. We assume the heating-up of the Universe to take place through the perturbative decay of the inflaton either into a pair of SM bosons or into a pair of SM fermions, while the inflaton itself oscillates in a monomial potential around the minimum.  In such a scenario the right handed neutrinos (RHN), that are responsible for asymmetry generation by their CP-violating decay, are produced both from the bath and from the perturbative two-body decay of the inflaton. Whereas, the DM is assumed to originate solely from the radiation bath. As a consequence, both the productions are controlled crucially by two factors: the inflaton coupling with the visible sector (bath), that decides the evolution of bath temperature during reheating and the shape of the potential during reheating. The latter also decides the background equation of state during the reheating era. We find, satisfying the cosmological bounds on the reheating temperature, it is possible to generate the observed baryon asymmetry [cf. Fig.~\ref{fig:asym}], along with adequate amount of DM during reheating with and without the explicit inflaton-RHN coupling. Interestingly, there exists a large DM mass window: MeV-PeV [cf. Fig.~\ref{fig:dm}], that can be produced via UV freeze-in during reheating. 

The current setup can leave its footprint at several low and high frequency gravitational wave (GW) detectors, thanks to the presence of the primordial inflationary GW spectrum which receives a blue tilt due to stiff period of expansion during reheating for certain types of monomial potential. One can relate the amount of blue-tilt to the inflaton-SM coupling, which, in turn is related to the generation of baryon asymmetry and DM [cf. Fig.~\ref{fig:pgw}]. Thus, the detection of such GW signal at future GW detectors can provide a test to the present scenario of high scale leptogenesis and non-thermal DM both of which lack direct detection prospects. To summarize, the present study provides a potential probe for scale of  leptogenesis (as well as DM mass scale) during reheating, that in turn is also a test for nonstandard cosmological history during the pre-BBN era.     
\section*{Acknowledgement}
BB would like to thank  Simon Cléry, Arghyajit Datta and Riajul Haque for fruitful discussions. The work of DB is supported by the Science and Engineering Research Board (SERB), Government of India grants MTR/2022/000575 and CRG/2022/000603. RR acknowledges financial support from the STFC Consolidated Grant ST/X000583/1.
\appendix
\section{Details of inflaton decay rate calculation}
\label{sec:inf-decay}
In a general monomial potential, the inflaton decay width can be parametrized as~\cite{Co:2020xaf,Ahmed:2021fvt,Barman:2022tzk,Barman:2023ktz}
\begin{align}\label{eq:decay-gen}
\Gamma_\phi = \Gamma_\phi^e \left( \frac{a_I}{a} \right)^\beta\,,
\end{align}
where $a_I$ is the initial value of the scale factor, indicating the end of inflation, $\Gamma_\phi^e$ denotes the inflaton width at $a=a_I$, while $\beta$ is assumed to be a constant parameter. Following~\cite{Barman:2023ktz}, the full expression for inflaton decay width reads
\begin{align}
\Gamma_{\phi \rightarrow NN} = \frac{1+n}{2n} \frac{1}{4\pi}  \left(\frac{y}{M_P} \right)^2\frac{\omega M_P^4}{\rho_\phi} \left(\frac{\rho_\phi}{\Lambda^4} \right)^{\frac{1}{n}}  \sum_{\ell=1}^\infty  \ell \lvert \mathcal{P}_\ell \rvert^2 (\ell \omega)^2\,,
\end{align}
where, the $\ell$ denotes the $\ell^{\rm th}$ oscillation modes of the inflaton field $\phi$.
For the T-model of inflation, the frequency $\omega$ is related to the effective mass of the inflaton field through
\begin{align}
\omega = m_\phi\,\sqrt{\frac{ \pi n}{2n-1}}\,\frac{\Gamma \left( \frac{n+1}{2n}\right)}{\Gamma \left( \frac{1}{2n}\right)}\,.
\end{align}
Its time-evolution is described by the power-law solution with the scale factor~\cite{Ahmed:2022tfm,Barman:2023ktz}
\begin{align}
&\omega = \omega_I \cdot  \left(\frac{\rho_\phi}{\Lambda^4}\right)^{\frac{n-1}{2n}} \simeq \omega_I \left( \frac{\rho_e}{\Lambda^4}\right)^{\frac{n-1}{2n}} \left(\frac{a_I}{a}\right)^{\frac{3(n-1)}{n+1}}, 
&\omega_I \equiv \sqrt{2\,n^2\,\pi }\, \frac{\Gamma \left( \frac{n+1}{2n}\right)}{\Gamma \left( \frac{1}{2n}\right)}\, \frac{\Lambda^2}{M_P}\,.
\end{align}
Gathering all the pieces we obtain
\begin{align}
\Gamma_{\phi\to \psi\psi}\simeq\omega_I\,\frac{1+n}{2n}\,\frac{y^2}{4\pi}\,  \left(\frac{\omega_I M_P}{\Lambda^2} \right)^2 \left(\frac{\rho_e}{\Lambda^4} \right)^{\frac{n-1}{2n}}  \sum_{\ell=1}^\infty\ell^3 \lvert \mathcal{P}_\ell\rvert^2\,.  
\end{align}
\begin{table}[t!]
\centering
$
\begin{array}{|c||c|c|c|c|c|c|c|c|c|c|c|}
\hline
n & 1 & 2 & 3 & 4 & 5  \\
\hline
\sum_l l |\mathcal{P}_l|^2 & 1/4 & 0.229 & 0.210 & 0.217 & 0.204  \\
\hline
\sum_l l^3 |\mathcal{P}_l|^2 & 1/4 & 0.240 & 0.244 & 0.250 & 0.257 \\
\hline
\end{array}
$\vspace{7pt}
\caption{Numerical sums to be used in Eq.~\eqref{eq:yeff} and \eqref{eq:mueff}.}
\vspace{7pt}
\label{tab:PkSums}
\end{table}
On comparison with Eq.~\eqref{eq:decay-gen},
\begin{align}\label{eq:yeff}
y_{\rm eff}^2 \equiv y^2 \sum_{\ell=1}^\infty\ell^3 \lvert \mathcal{P}_\ell\rvert^2\,,  
\end{align}
where the sum is given in Tab.~\ref{tab:PkSums}. For  bosonic decay, 
\begin{align}
\Gamma_{\phi \rightarrow \varphi\varphi} &= \frac{1+n}{2n} \frac{\omega }{2\pi}  \left(\frac{\mu\,M_P}{\Lambda^2} \right)^2 \left(\frac{\rho_\phi}{\Lambda^4} \right)^{\frac{1-n}{n}}\sum_{l=1}^\infty  l \lvert \mathcal{P}_l \rvert^2\,,    
\end{align}
implying
\begin{align}\label{eq:mueff}
\mu_{\rm eff}^2\equiv\mu^2\,\sum_{l=1}^\infty\ell\lvert \mathcal{P}_\ell \rvert^2    
\end{align}

\bibliography{ref}
\bibliographystyle{JHEP}
\end{document}